\documentclass[aps,prx,10pt,amssymb,amsmath,superscriptaddress,tightenlines,twocolumn,notitlepage,
] {revtex4-2}
\usepackage{graphicx}
\usepackage{amsmath}
\usepackage{amssymb}
\usepackage{bm}
\usepackage{color}
\usepackage{hyperref}
\usepackage{tabularx}
\usepackage{multirow}
\usepackage{booktabs}
\usepackage{braket}
\usepackage{epstopdf}

\begin{document}

\title{Multiple Chern bands in twisted MoTe$_2$ and possible non-Abelian states}

\author{Cheng Xu}
\affiliation{Department of Physics and Astronomy, University of Tennessee, Knoxville, TN 37996, USA}
\affiliation{Department of Physics, Tsinghua University, Beijing 100084, China}

\author{Ning Mao}
\affiliation{Max Planck Institute for Chemical Physics of Solids, 01187, Dresden, Germany}

\author{Tiansheng Zeng}
\affiliation{Department of Physics, Xiamen University, China}

\author{Yang Zhang}
\affiliation{Department of Physics and Astronomy, University of Tennessee, Knoxville, TN 37996, USA}
\affiliation{Min H. Kao Department of Electrical Engineering and Computer Science, University of Tennessee, Knoxville, Tennessee 37996, USA}

\begin{abstract}

We investigate the moir\'e band structures and possible even denominator fractional quantum Hall state in small angle twisted bilayer MoTe$_2$, using combined large-scale local basis density functional theory calculation and continuum model exact diagonalization. Via large-scale first principles calculations at $\theta=1.89^{\circ}$, we find a sequence of $C=1$(Chern number in K valley)moir\'e Chern bands, in analogy to Landau levels. 
By constructing the continuum model with multiple Chern bands, we undertake band-projected exact diagonalization using unscreened Coulomb repulsion to identify possible non-Abelian states  near twist angle $\theta=1.89^{\circ}$ at the half filling of second moir\'e band.

\end{abstract}

\maketitle


Moir\'e materials based on transition metal dichalcogenides (TMDs) have emerged as a promising domain for exploring novel quantum phenomena \cite{kennes2021moire,mak2022semiconductor}. Owning to the substantial effective mass inherent to TMDs valence bands and the persistence of narrow moir\'e bands across various twist angles, these materials showcase a diverse array of correlated electron states, characterized by pronounced interaction effects, including Mott and charge-transfer insulators\cite{PhysRevLett.121.026402,PhysRevB.102.201115,regan2020mott,tang2019wse2,wang2020correlated,ghiotto2021quantum,li2021continuous,https://doi.org/10.48550/arxiv.2202.02055}, generalized Wigner crystals \cite{regan2020mott,xu2020correlated,zhou2021bilayer,jin2021stripe,li2021imaging,huang2021correlated,padhi2021generalized,matty2022melting}, and quantum anomalous Hall (QAH) effect \cite{li2021quantum}. Remarkably, recent transport experiments on twisted MoTe$_2$ have provided unambiguous evidences of both integer and odd-denominator fractional quantum anomalous Hall (FQAH) effects\cite{park2023observation,xu2023observation}, and the signature for fractional quantum spin Hall effect at hole filling factor $\nu=-3$ \cite{kang2024evidence}. The observations of FQAH were made within a range of fairly large twist angles, specifically $\theta\sim 2.7^{\circ}-3.9^{\circ}$, evidenced through both optical \cite{cai2023signatures} and compressibility \cite{zeng2023thermodynamic} measurements, within the first moir\'e valence band. The recent experiment on fractional quantum spin Hall (FQSH) effect \cite{kang2024evidence} at $\theta\sim 2.1^{\circ}$ revealed the remarkable triple quantum spin Hall effects, driving the interest to higher filling factors and small twisted angles. 

The realization of fractional quantum anomalous Hall effect not only fundamentally broadens the taxonomy of topological phases of matter but also holds promising prospects for harnessing the power of anyons in topological quantum computations at zero magnetic field \cite{parameswaran2013fractional,bergholtz2013topological,neupert2015fractional,liu_recent_2023,RevModPhys.80.1083}. On the theory side, the fractional Chern insulator (FCI) phases in topological flat band systems has been proposed for over a decade \cite{PhysRevLett.106.236802,sheng2011fractional,regnault2011fractional,sun2011nearly,sun2011nearly}. In recent years, within the graphene \cite{repellin2020chern,ledwith2020fractional,wilhelm2021interplay,parker2021field}  and
TMD \cite{devakul2021magic,li2021spontaneous}-based moir\'e system, theoretical predictions have pointed to such an exotic state at partial filling of the topological moir\'e flat band at long moir\'e wavelength. 
Non-Abelian quantum Hall states such as the possible Moore-Read state~\cite{MOORE1991362} at even-denominator filling factor $\nu=5/2$ has been discussed in Landau level~\cite{PhysRevLett.59.1776}. Under particle-hole symmetry breaking~\cite{PhysRevLett.99.236806,PhysRevLett.99.236807} (e.g. Landau level mixing or local three-body interaction), several numerical studies~\cite{PhysRevLett.101.016807,PhysRevLett.104.076803,PhysRevLett.106.116801,PhysRevX.5.021004} have shown that the Moore-Read Pfaffian (anti-Pfaffian) state with six-fold ground state degeneracies may be favored. Numerical explorations have also suggested that topological flat band models may host a fermionic non-Abelian Moore-Read state under synthetic three-body interaction~\cite{PhysRevB.85.075128,zhang2024moore} or long range dipolar interaction~\cite{PhysRevB.91.125138}. However, a realistic simulation of such a non-Abelian state in a microscopic lattice model remains challenging. Recent experiment on fractional quantum spin Hall effect~\cite{kang2024evidence} highlights the possibility of a time-reversal pair of even-denominator $3/2$ FQAH states, offering a strong candidate for a non-Abelian state within twisted MoTe$_2$'s topological minibands. Motivated by these developments, we theoretically analyze and propose the realistic models for its emergence.

In this work, we start from the local basis first-principle calculations and continuum model at $\theta=1.89^{\circ}$, and study the possible non-Abelian state at half filling of the second moir\'e valence band. From density functional calculations (DFT), we find the number of $C=1$ Chern bands increases from 2 \cite{reddy2023fractional,xu2024maximally,zhang2023polarization} to 5 when twist angle changes from $\theta=5.08^{\circ}$ to $\theta=1.89^{\circ}$. The Chern numbers of DFT bands are directly calculated from Berry curvature integral and Wilson loop from Kohn-Sham wavefunctions. Moreover, we confirm the multiple $C=1$ bands from edge states calculations.

For the fitting of the continuum model at $\theta=1.89^{\circ}$, we constrain the parameter space via fixing the Chern number of top five valence bands and increasing the weight of second moir\'e bands. With the small angle continuum model, we find strong evidences for the Moore-Read states through momentum-space exact diagonalization spectrum in several lattice geometries. The ground state degeneracy and momentum locations fulfill the generalized Pauli principle \cite{Haldane1991b,Bernevig2012} for Moore-Read state ~\cite{MOORE1991362} for even and odd number of electrons, similar to those in the half-filled first Landau level. 
Our work provides a comprehensive analysis of multiple topological bands in twisted MoTe$_2$ and the continuum model for realizing non-Abelian states, paving the way for its material realizations.


\begin{figure}
\includegraphics[width=\columnwidth]{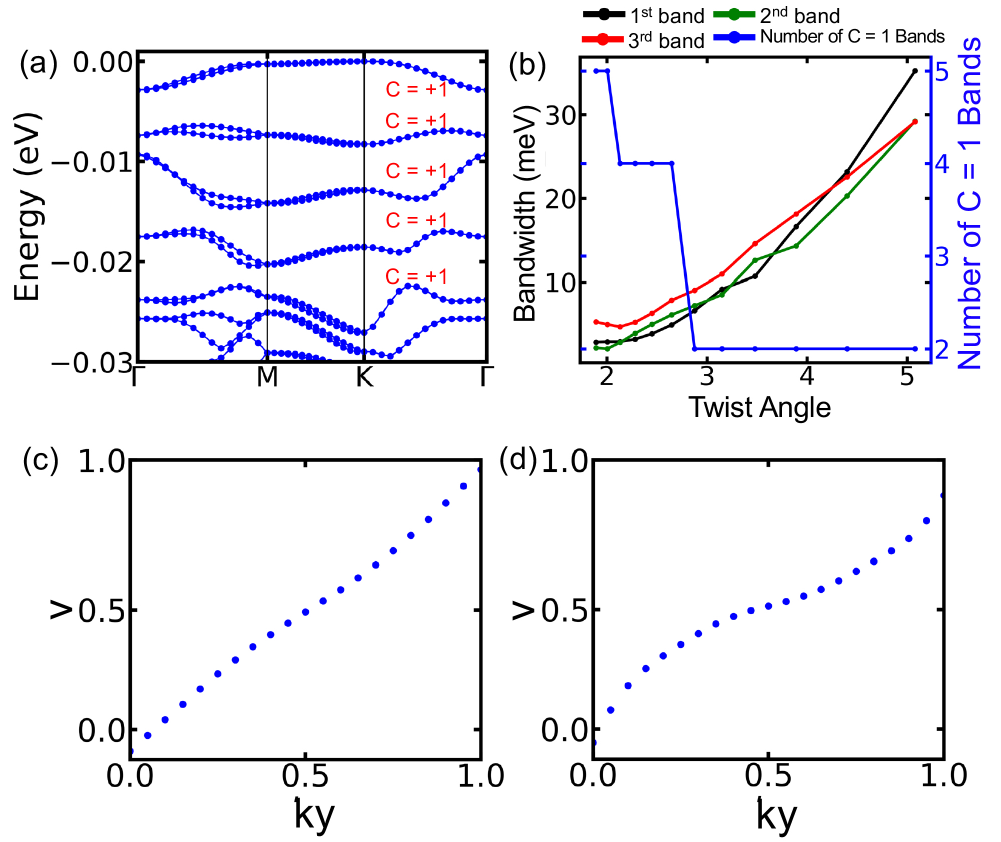}
\caption{(a) Band structures for homobilayer MoTe$_2$ with
twist angle 1.89$^\circ$, where the five flat bands near the Fermi level possess a Chern number of $C = 1$. (b) Angle dependent band width and number of $C=1$ bands 1.89$^\circ$ to 5.08$^\circ$, calculated by local-basis DFT. The black, green, and red dotted lines denote for the band width of the first, second, and third band. The blue dotted line represent for the number of C = 1 bands. The evolution of Wannier charge center for the (c) first band and (d) second band, indicating that C = 1.}\label{band}
\end{figure}

\textbf{DFT results.} 
 Here we employ the local-basis OpenMX package on the twist angle $\theta=1.89^\circ$. There are 1838 Mo atoms and 3676 Te atoms, forming a Hamiltonian with 191152 dimensions \cite{mao2023lattice}. Since full diagonalization of such a Hamiltonian is unrealistic in present hardware platform, we apply the shift-and-invert trick, recasting the generalized eigenvalues problem to the eigenvalues problem, and then apply the Lanczos algorithm to get part of the eigenvalues near Fermi level as commonly used in linear-scaling DFT~\cite{bowler2012methods,zhang2020density}.

As shown in Fig.~\ref{band}, our DFT calculations demonstrate a clear trend: a decrease in the twist angle leads to a reduction in the bandwidth near the Fermi level. For a relatively large twist angle of 5.08$^\circ$, the smallest bandwidth is 29 meV for the second band, consistent with plane-wave DFT \cite{xu2024maximally,mao2023lattice}. However, at the small twist angle of 1.89$^\circ$, five nearly flat bands emerge near the Fermi level, with a small bandwidth of 2.9, 2.2, 5.3, 3.7, and 4.6 meV, respectively. As a result of $C_{2y}$ symmetry, bands along the $\Gamma K$ lines are doubly degenerate, while a clear splitting is observed along the $\Gamma M$ line as shown in Fig.~\ref{band}(a). Furthermore, our results show multiple band inversions near a twist angle of 1.89$^\circ$, leading to the formation of topological flat bands. These bands are distinctively separated from others by a band gap ranging from 1.9 meV to 3.9 meV. The band separation hints at the possibility of various QSH states \cite{kang2024evidence}, including single, double, triple QSH states, with increased hole doping in the system.

With twisted Hamiltonian, we then proceed to study the topological properties of narrow bandwidth moir\'e valence band. It is worth noting that the previous $C_3$ symmetry indicator at $C_{3z}$ symmetric momenta fails to accurately determine Chern numbers with modulo three \cite{zhang2021spin}. Achieving precise calculations necessitates the integration of Berry curvature or Berry connection across momentum space. Despite Kramer pairs have a total Chern number of zero due to time-reversal symmetry, a significant Berry curvature can still arise within a single valley. To isolate the single valley bands within the Hamiltonian, we construct the valley operator $\hat{s}_{z}$ based on the atomic positions and orbital elements, and separate the single-valley eigenvectors $\left | u_{m}(\mathbf{k}) \right\rangle$ from the entire eigenvectors $\left | v_{m}(\mathbf{k}) \right\rangle$ based on the expectation value of $\hat{s}_{z}$:
\begin{equation}
    S_{mn }^{z}(\mathbf{k})=\left\langle u_{m}(\mathbf{k}) 
 |\hat{s}_{z}|  u_{n}(\mathbf{k})\right\rangle
\end{equation}
Across the two-dimensional Brillouin zone, most of the lines—excluding the $\Gamma K$ and $M K$ lines—are non-degenerate, with their expectation values of $\hat{s}_{z}$ approaching either $+1$ or $-1$. For those lines that are degenerate, a basis transformation is implemented to construct the disentangled wavefunctions, characterized by expectation values of $+1$ or $-1$. This allows us to feasibly separate the single-valley eigenvectors from the valley-mixed eigenvectors.

Therefore, it is practical to calculate the Chern number for the single band  from the eigenvector in each valley. Due to the large number of atomic orbitals, the Kubo formula approach requiring full diagonalization is inapplicable here. Therefore, we calculate the Chern number through the Fukui-Hatsugai-Suzuki method~\cite{fukui2005chern}:
\begin{equation}
\begin{aligned}
&U_{\Delta \mathbf{k}}(\mathbf{k})=\frac{\left\langle u(\mathbf{k}) \mid u(\mathbf{k}+\Delta \mathbf{k})\right\rangle}{|\left\langle u(\mathbf{k}) \mid u(\mathbf{k}+\Delta \mathbf{k})\right\rangle|} \\
    &F(\mathbf{k})=\operatorname{Im} \log U_{\Delta\bm{k_{1}}}(\mathbf{k}) U_{\Delta\bm{ k_{2}}}\left(\mathbf{k}+\Delta \bm{k_{1}}\right) \times \\ 
    &U_{\Delta \bm{k_{1}}}^{-1}\left(\mathbf{k}+\Delta\bm{ k_{2}}\right) U_{\Delta \bm{k_{2}}}^{-1}(\mathbf{k}), 
\end{aligned}
\end{equation}
where $ U_{\Delta \bm{k_{1}}}(\mathbf{k})$ 
is the U(1) link variable from the single-particle wave function $u(\bm{k})$, and $F(\bm{k})$ is the lattice field strength, which is related to the Chern number as:$C=\frac{1}{2 \pi}  \sum_{\bm{k}} F (\mathbf{k}) $. To further verify the band topology, we also examine the evolution of Wannier charge centers(WCC)~\cite{wcc1,wcc2}:
\begin{equation}
    \nu_{n} (k_y)= \int \left\langle u_{n}(\mathbf{k})\left|\partial_{k_{x}}\right| u_{n}(\mathbf{k})\right\rangle d k_{x}
\end{equation}
As illustrated in Fig.~\ref{band}(c) and ~\ref{band}(d), the number of crossings between the WCC and any horizontal lines
is 1, indicating the Chern number of 1. 
Furthermore, $C=1$ for top five moir\'e bands will lead to the presence of multiple pairs of gapless edge states inside bulk gap. This is clearly illustrated in Fig.S1(b), confirming the topological properties of the flat Chern bands.

\textbf{Continuum model with high harmonic term.}
\label{continuummodel}
To perform the many-body calculation, we start with the continuum model for twisted MoTe$_2$ \cite{wu_topological_2019}. Here we incorporate the higher-order harmonic terms for both inter-layer and intra-layer coupling, extending up to the second harmonics. Considering the significant momentum difference between the two valleys ($1/a\gg 1/a_m$), we ignore the inter-valley coupling, focusing solely on the K valley. Additionally, we limit our model to single-spin, owning to the large Ising spin-orbit coupling in MoTe$_2$. As a result, we arrive at the following continuum model for the K valley:
\begin{equation}
\hat{H_s}=	\begin{bmatrix}
		-\frac{\boldsymbol{(\hat k-K_{t})^2}}{2m^*}+\Delta_{t}(\boldsymbol{r})& \Delta_{T}(\boldsymbol{r})\\
		\Delta^{\dagger}_{T}(\boldsymbol{r}) & -\frac{\boldsymbol{(\hat k-K_{b})^2}}{2m^*}+\Delta_{b}(\boldsymbol{r})
	\end{bmatrix}	
\end{equation}
with:
\begin{equation}
	\begin{aligned}
		\Delta_{l}(\boldsymbol{r})&=2V_1\sum_{i=1,3,5}\cos(\boldsymbol{g^{1}_i\cdot r}+l\phi_1)+2V_2\sum_{i=1,3,5}\cos(\boldsymbol{g^{2}_i\cdot r})\\
		\Delta_{T}&=w_1\sum_{i=1,2,3}e^{-i\boldsymbol{q^1_i\cdot r}}+w_2\sum_{i=1,2,3}e^{-i\boldsymbol{q^2_i\cdot r}}\\
	\end{aligned}
\end{equation}
where $\boldsymbol{\hat{k}}$ is the momentum operator, $\boldsymbol{K_{t}}(\boldsymbol{K_{b}})$ is high symmetry momentum $\boldsymbol{K}$ of the top(bottom) layer, $\Delta_{t}(\boldsymbol{r})(\Delta_{b}(\boldsymbol{r}))$ is the layer dependent moir\'e potential, $l=+1 $ for top layer and $l=-1 $ for bottom layer; $\Delta_{T}(\boldsymbol{r})$ is the interlayer tunneling, $\boldsymbol{G_i}$ is moir\'e reciprocal vector, $\boldsymbol{g_i^1}$ and $\boldsymbol{g_i^2}$ represent the momentum differences between the nearest and second-nearest plane wave bases within the same layer. Similarly, $\boldsymbol{q_i^1}$ and $\boldsymbol{q_i^2}$ denote the momentum differences between the nearest and second-nearest plane wave bases across different layers ( a detail description of these vectors is shown in the Fig.~S14). And in this continuum model, there is obviously an effective inversion symmetry, which makes the band is always double degenerate. This symmetry is artificial because we only retain the amplitude of the second harmonic term. However, since the bands from DFT in both valleys are nearly degenerate,  this method is appropriate.

\begin{figure}[t]
\includegraphics[width=1\columnwidth]{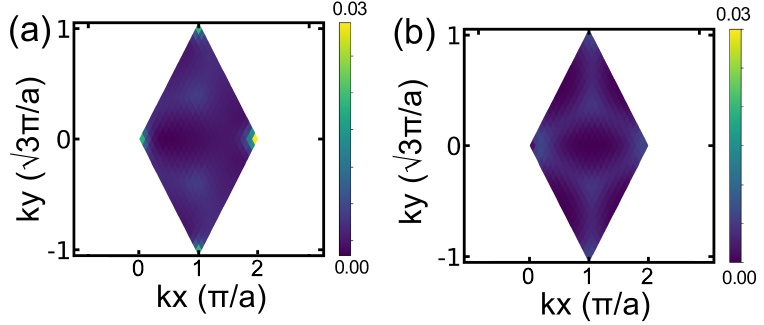}
\caption{The distributions of $F(\bm{k})$(the integral of the Berry curvature around the small loop in the $\bm{k}$-mesh.) across the two-dimensional Brillouin zone for the second band, calculated by (a) first-principles calculations and (b) continuum model. We note the distribution is similarly positive and uniform, and sums to a Chern number of 1. a is the lattice constant of twisted MoTe$_2$ at 1.89$^\circ$.
}\label{chern}
\end{figure}

To derive the parameters of continuum model, we adopt two guiding principles in the fitting of DFT band structures. Firstly, we ensure that the Chern numbers for top five valence bands are equal to 1, closely related to Landau levels \cite{morales2024magic}. Subsequently, particular focus is placed on the second band, for which we minimize the error of band dispersion and Berry curvature distribution as much as possible. From our DFT calculation at $\theta=1.89^\circ$, we obtain the parameters:
\begin{equation}
	\begin{aligned}
		&\phi_1=-90.0^\circ,V_1=2.4 \ meV,V_2 = 1.0\  meV\\
		 &w_1=-5.8 \ meV, w_2 =2.8\ meV
	\end{aligned}
\end{equation}

The fitting results are shown in Fig. S1(a), where the major topological features are captured in the continuum model. Figure ~\ref{chern}(b) displays the distributions of Berry curvature, where both the shape and magnitude closely resemble those obtained from DFT calculations. 
Additionally, the new parameters effectively capture the variation in the Chern number with the twist angle, as demonstrated in Fig. S17(b), where the top three bands consistently exhibit $C = 1$ within the range of $1^\circ$ to $3^\circ$.
It is found that the potential terms are significantly smaller than those derived from the DFT bands at larger twist angle $\theta>2.89^\circ$ \cite{xu2024maximally,mao2023lattice,jia2023moir,wang2024fractional}. This difference indicates that the parameters previously employed do not effectively describe the higher moir\'e valence bands at small twist angles ($\le 2.5 ^\circ$). 

\begin{figure}[t]
\includegraphics[width=1\columnwidth]{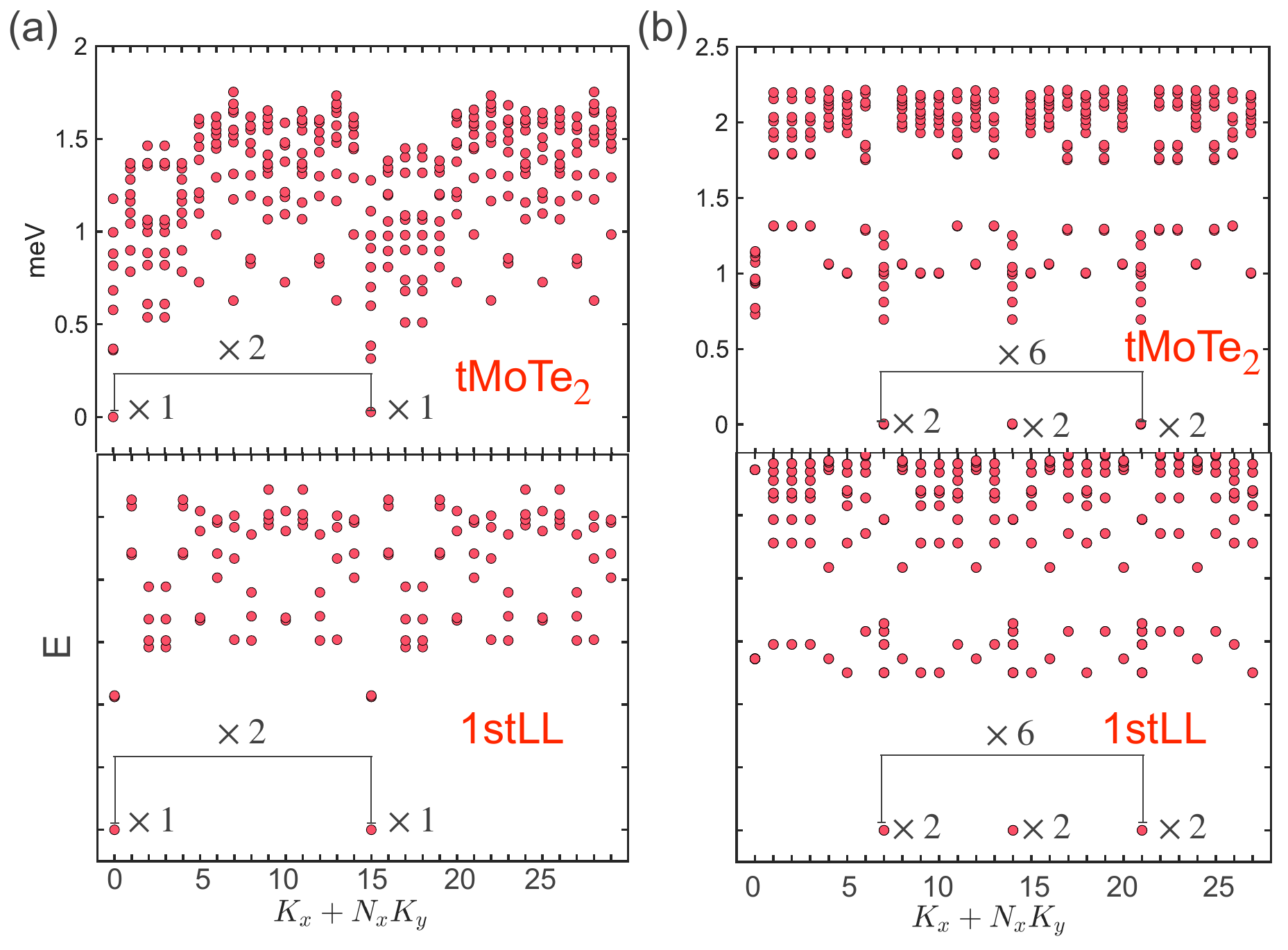}
\caption{Exact diagonalization spectrum. (a)/(b)(top). The many-body spectrum from the ED calculation on the 30/28 sites cluster at half filling of second moir\'e band with SCHF renormalized bands at $\theta = 1.6^\circ$ and dielectric constant $\epsilon = 5$. (a)/(b)(bottom). ED spectrum of the 1st Landau Level. The ground state manifold agrees well with our results from continuum model.
}\label{ED}
\end{figure}
\begin{figure}[t]
\includegraphics[width=1\columnwidth]{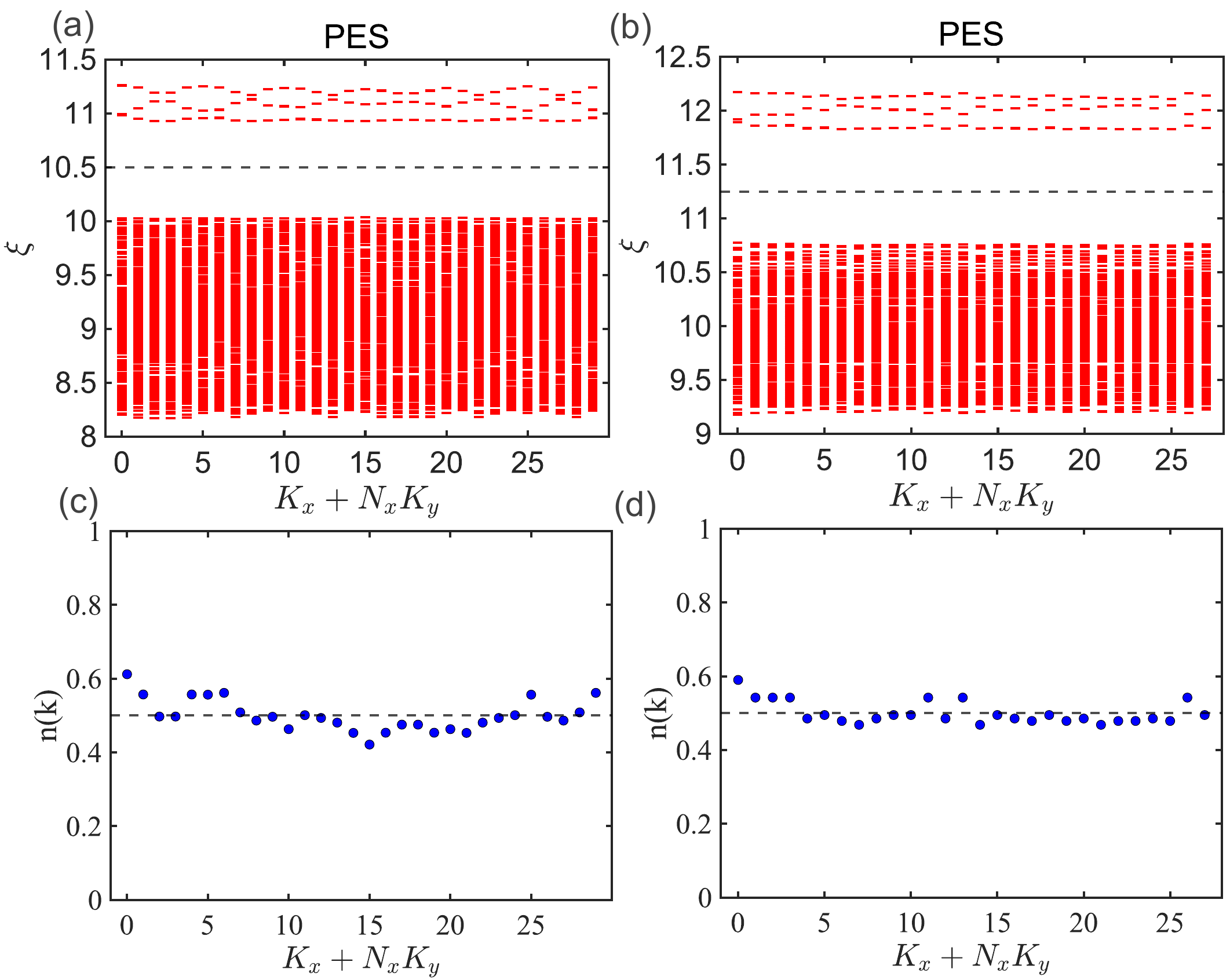}
\caption{(a).The particle entanglement spectrum calculated on the 30-site cluster by the dividing the system into $N_A=3$ and $N_B=N_e-N_A$. The number of states below the gap (the black dashed line) is 3970, which matches with the counting of quasihole excitation of MR states. (b). The same as (a) on the 28-site cluster with cutting corresponding  to $N_A=3$ and there are 3192 states below the gap. (c)/(d). The momentum distribution function $n(\bm{k})=\frac{1}{N_{gs}}\sum_{i}\langle \Psi_{i}|\hat{n}_{\bm{k}}|\Psi_i\rangle $ of 30-site and 28-site clusters, respectively. The dashed line represents the average value of $n(\bm{k})$. 
}\label{NFig4}
\end{figure}

\textbf{Exact Diagonalization at half filling of 2nd moir\'e band.}

In this section, we focus on the half filling of 2nd moir\'e band. To account for the effect of the filled first moiré band, we perform a self-consistent Hartree-Fock calculation(SCHF) at $\nu = -2$ . We then project the Coulomb interaction onto the second moiré band and begin with the assumption of spontaneous spin polarization to reduce the Hilbert space dimension. The complete many-body Hamiltonian is then expressed as:
\begin{equation}
	\begin{aligned}		
		H&=H_s+V,\\
V&=\sum_{s,s^\prime}\frac{1}{2}\int\int d\boldsymbol{r_1}d\boldsymbol{r_2} V(\bm{r_1-r_2})\hat{c}_{s}^{\dagger}(\boldsymbol{r_1})\hat{c}_{s^\prime}^{\dagger}(\boldsymbol{r_2})c_{s^\prime}(\boldsymbol{r_2})c_{s}(\boldsymbol{r_1})
	\end{aligned}
\end{equation}
Here, $H_s$ is the single particle continuum model Hamiltonian of hole and V is the Coulomb interaction, $c^\dagger_{s}(\bm{r})$ is the creation operator of hole in real space, $s$ is the spin index and we use long range Coulomb interaction:$V(\bm{r_1-r_2})=\frac{e^2}{\epsilon |\bm{r_1}-\bm{r_2}|}$ and choose the realistic dielectric constant $\epsilon = 5 $.
We then project the model Hamiltonian into the second moir\'e valence bands:
\begin{equation}
\begin{aligned}
&	H_s =\sum_{nks}\epsilon _{nks}a^\dagger_{nks}a_{nks},\\
&	V=\frac{1}{2}\sum_{\bm{n}\bm{s}\bm{k}}V_{n_1k_1n_2k_2n_3k_3n_4k_4}^{s_1s_2s_3s_4}\hat{a}^\dagger_{n_1k_1s_1}	\hat{a}^\dagger_{n_2k_2s_2}\hat{a}_{n_3k_3s_3}\hat{a}_{n_4k_4s_4},\\	
&	V_{n_1k_1n_2k_2n_3k_3n_4k_4}^{s_1s_2s_3s_4}=\int\int d\boldsymbol{r_1}d\boldsymbol{r_2} V(\bm{r_1-r_2})\times \\
&\psi^*_{n_1k_1s_1}(\bm{r_1})\psi^*_{n_2k_2s_2}(\bm{r_2})\psi_{n_3k_3s_3}(\bm{r_2})\psi_{n_4k_4s_4}(\bm{r_1})\label{EDHam}
\end{aligned}
\end{equation}
where n is the band index, $a^\dagger_{nks}$ is the creation operator of Bloch states, $\psi_{nks}$ is the eigen-vector of continuum model with SCHF, $\hat{c}^{\dagger}_{s}(\bm{r})=\sum_{nks}\psi^*_{nks}(\bm{r})\hat{a}^{\dagger}_{nks}	$, $\sum_{\bm{n}} =\sum_{n_1n_2n_3n_4}$ and so on. 

Our calculations focus on the half filling of second moir\'e valence band around the DFT twisted angle $1.89^\circ$, and we identify most stable Non-Abelian states near $1.60^\circ$ from various numerical evidences \cite{kang2024evidence}. Our primary findings are depicted in Fig.~\ref{ED}, revealing several quasi-degenerate ground states. Specifically, we identify two-fold quasi-degenerate states for the 30-site cluster (15 electrons),and six-fold degenerate states for the 28-sites cluster (14 electrons). Both the momentum location and ground state degeneracy agree well with the requirement of the generalized Pauli principle, where no more than two particles in four consecutive orbitals, called (2,2)-admissible ``root'' configuration\cite{Haldane1991b,Bernevig2012} . In short, for an odd number of electrons, the possible (2,2)-admissible``root'' configuration~\cite{PhysRevLett.100.246802} requires that only the occupation partition ``$1010\cdots101010$" and its translational invariant partner``$0101\cdots010101$" are allowed, explaining the two-fold degeneracy. But for even number of electrons, there are six possible configurations, which explain the six-fold degeneracy as demonstrated in the Supplemental Material. The distinct degeneracies observed for odd and even number of electrons provide compelling evidence that the ground states may be the Non-Abelian Moore-Read states.

Moreover, we perform a comparative examination of the states within twisted MoTe$_2$ and the 1st Landau Level (1st LL), uncovering notable parallels on the ground states, as depicted in Fig.~\ref{ED}. Additionally, we extend our calculations to clusters of varying sizes, with detailed methodologies outlined in the Supplemental Material. Together, these findings underscore the remarkable similarity to the 1st LL across different system configurations.

We further calculate the particle entanglement spectrum within the ground state manifold. By dividing the system into $N_A$ and $N_e - N_A$, we observe a visible gap in the spectrum, with the number of states below the gap corresponding to the quasihole excitations, as dictated by the generalized Pauli principle. Notably, this result remains stable across various system sizes—not only for the 28-site and 30-site systems shown in Fig.~\ref{NFig4}, but also for 4x6-site, 3x8-site, and 26-site clusters, as detailed in the supplementary materials. Additionally, we present the momentum distribution function $n(\bm{k})$ in Figs.~\ref{NFig4}(a) and~\ref{NFig4}(b), which is nearly uniform, effectively ruling out the possibility of a charge density wave state. To further confirm the topology of the ground states, we computed the many-body Chern numbers (see the Fig. S11 in the supplementary material) for both 12-site and 28-site systems, which give a perfect quantization of $C = 1/2$ per state. In summary, our calculations provide strong evidence supporting the existence of potential non-Abelian states.

Given the preservation of particle-hole symmetry in dispersionless Landau levels, the ground states are expected to be symmetrized Moore-Read states, comprising superpositions of Pfaffian and anti-Pfaffian states, owing to potential finite size effects \cite{PhysRevLett.101.016807,PhysRevB.80.241311,PhysRevB.95.201116}. However, the missing of particle-hole symmetry in our moir\'e band-projected Hamiltonian would restrict the non-Abelian nature of the ground states at half filling of second moir\'e bands. Consequently, further investigation is warranted to elucidate the precise nature of the topological order in this context.


In this work, we investigate the single-particle electronic structures of small-angle moir\'e MoTe$_2$ with local basis DFT, exploring its topological properties through Berry curvature and Wilson loop calculations. Specifically, we concentrate on the second moir\'e band, constructing a continuum model comprising up to five $C=1$ Chern bands. This model offers insight into various charge fractionalization phenomena reminiscent of those observed in Landau level systems.

Drawing an analogy to the first excited Landau level, our exact diagonalization reveals a remarkably similar many-body spectrum  in both even and odd electron systems, providing compelling evidences for the existence of non-Abelian states. We note that in realistic situations, breaking the particle-hole symmetry will lead to the favored Moore-Read Pfaffian. Numerically, the Pfaffian or antiPfaffian nature of the ground states may be distinguished by adding opposite three body interactions, or density-matrix renormalization group calculation of their entanglement spectrum, originating from the different edge structures. 

The discovery of integer and fractional quantum Hall effects in the moir\'e MoTe$_2$ \cite{park2023observation,xu2023observation} and pentalayer graphene \cite{lu2024fractional} at zero magnetic field provides ideal material platforms for the realization of charge fractionalization beyond the conventional two-dimensional electron gas at high magnetic field. Similar to a partially filled Landau level, a partially filled topological band can exhibit a symphony of distinct phases as a function of filling factor, each bringing its own novelty as an impetus to extend the frontier of condensed matter physics. While earlier investigations primarily centered on first Chern bands, our study broadens this focus to higher moir\'e bands, uncovering a series of $C=1$ bands, laying the foundation for the study of higher filling factor fractional states \cite{kang2024evidence}.

{\it Note: Near the completion of this work, we became aware of a related work~\cite{reddy2024non}, which studied the non-Abelian state using Skyrmion model and its application in twisted semiconductor bilayers.}

\section*{Acknowledgments}
We are grateful to Kin Fai Mak, Ahmed Abouelkomsan, Liang Fu, Kai Sun, Zhao Liu for their helpful discussions. Y. Z. is supported by the start-up fund at University of Tennessee Knoxville.
\bibliography{moire_MoTe}

\clearpage
\pagebreak
\onecolumngrid

\begin{center}
    \textbf{Supplementary material for: Multiple Chern bands in twisted MoTe$_2$ and possible non-Abelian states}
\end{center}
\setcounter{figure}{0}
\renewcommand{\thefigure}{S\arabic{figure}}
\setcounter{equation}{0}
\renewcommand{\theequation}{S\arabic{equation}}
\setcounter{table}{0}
\renewcommand{\thetable}{S\arabic{table}}

\subsection{Lattice relaxation for homobilayer MoTe$_2$ with twist angle $1.89^{\circ}$}

To relax the structure, we utilize the ab initio deep potential (DP) molecular dynamics method. Our methodology begin with constructing $3 \times 3 \times 1$ MM, MX, and XM configurations and 28 distinct intermediate transition states~\cite{zhang2018deep}. Each configurations are relaxed at a fixed volume and then subjected to 200 random perturbations. We then gather initial data sets through 20 fs ab initio molecular dynamics simulations, calculating energies, forces, and virial tensors using VASP. This data set is used to train the DP model, utilizing a descriptor (DeepPot-SE) for both angular and radial atomic configurations and embedding layers mapping descriptors to atomic energies. Following initial training, the model undergoes molecular dynamics simulations across various pressures and temperatures, generating trajectories categorized based on model deviation. Selected configurations undergo self-consistent density functional theory calculations for further training iterations. Furthermore, we expanded our training data with large-angle twisted structures and applied transfer learning principles. By freezing embedding layer parameters and focusing on the hidden and output layers, we 
construct the transform learning neural network that can be used to relax the homobilayer MoTe$_2$ with twist angle 1.89$^{\circ}$.
\begin{figure}[h]
\includegraphics[width=0.8\columnwidth]{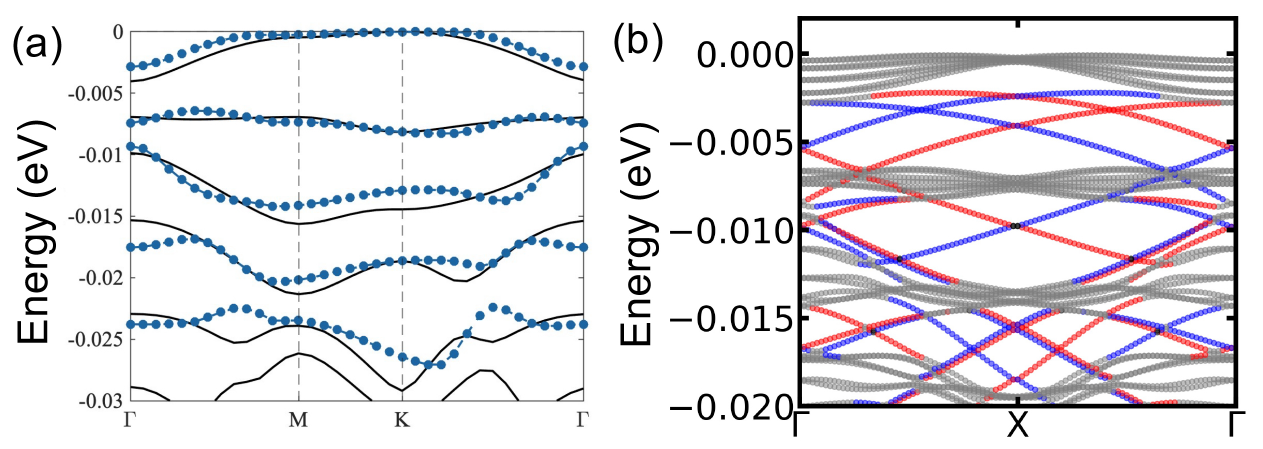}
\caption{(a) The comparative analysis of the band structures for twist angle 1.89$^{\circ}$. Blue points/lines illustrate the results from OpenMX calculations, while the black line represents the fitting results from continuum model. (b) Band structure of twisted MoTe$_2$ nanoribbon, calculated from OpenMX hamiltonian. The trivial bulk states are shown in gray, while spin-up/-down edge states are in red/blue.}
\label{SM_band_edge}
\end{figure}

\begin{figure}[h]
\includegraphics[width=0.8\columnwidth]{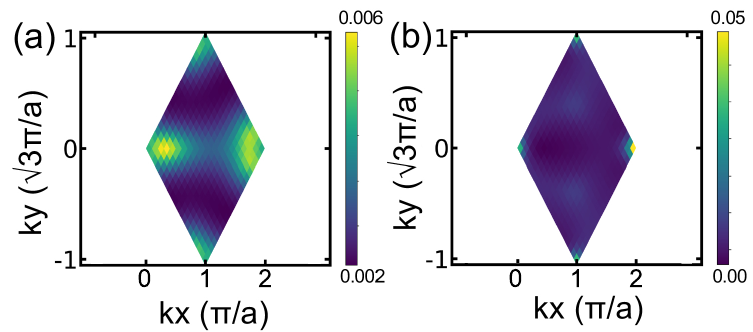}
s\caption{The DFT Berry curvature distributions of $\theta=1.89^{\circ}$ for the (a) top valence band and (b) second valece band.}
\label{SM_topology}
\end{figure}
\textbf{Details about openMX calculations.} Our openMX basis are Projected Atomic Orbitals (PAOs) specified as \textit{Mo7.0-s3p2d1} and \textit{Te7.0-s3p2d2}~\cite{openmx_basis,openmx_largescale}. The notation \textit{7.0} indicates a cutoff radius of 7.0 Bohr. For \textit{Mo7.0-s3p2d1}, \textit{s3p2d1} denotes the inclusion of 3 sets of $s$-orbitals, 2 sets of $p$-orbitals, and 1 set of $d$-orbitals, totaling 14 atomic orbitals. Similarly, \textit{Te7.0-s3p2d2} includes 3 sets of $s$-orbitals, 2 sets of $p$-orbitals, and 2 sets of $d$-orbitals, amounting to 19 atomic orbitals. These configurations are used to perform the self-consistent calculations. 

\subsection{Chern number of top two band}

Figure.~\ref{SM_topology}(a) and ~\ref{SM_topology}(b) present the distribution of Berry curvature. The Chern numbers, derived from the integral of Berry curvature, are calculated to be 1 for both the first and second band. Additionally, gapless edge states are observed within the gap between the first and second band, as well as between the second and third band. This phenomenon is explicitly depicted in Fig.~\ref{SM_band_edge}(b), thereby affirming the nature of multiple Chern numbers within separate valley.

\subsection{The momentum geometry}
\begin{figure}[h]
\includegraphics[width=\columnwidth]{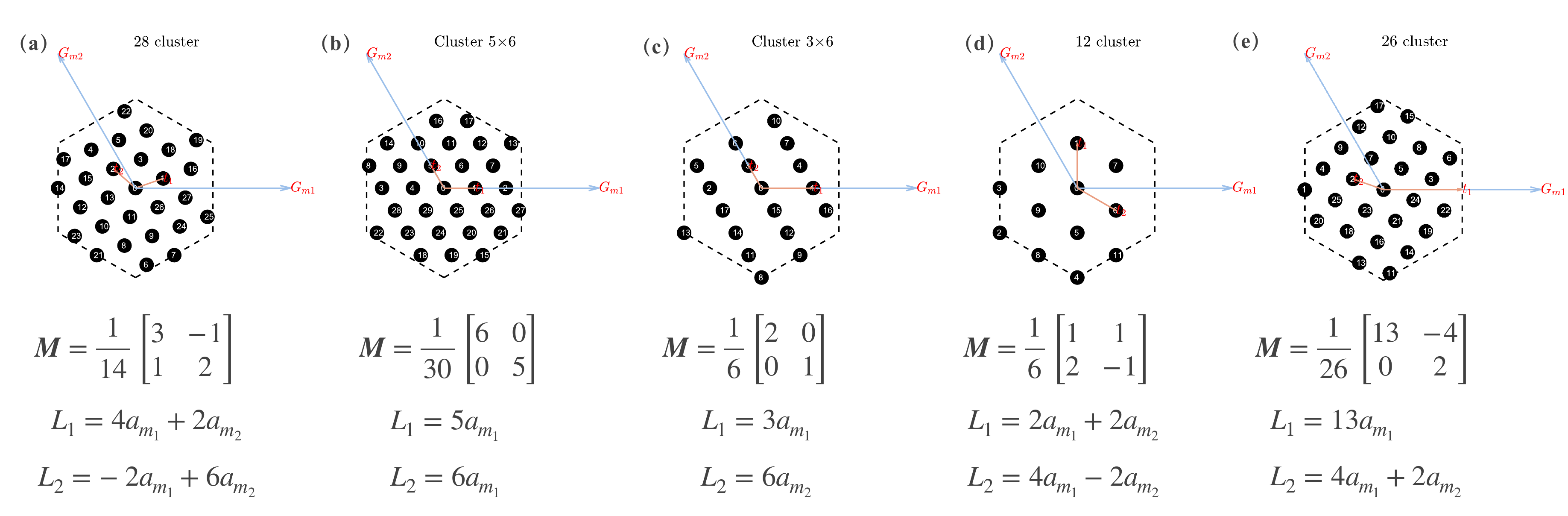}
\caption{The finite-sized cluster we use in the ED calculation. The $L_1,L_2$ are the lattice vector of the real-space cluster, $G_{m1},G_{m2}$ are the reciprocal vector of moire system. M is the matrix which defines the relation between the $t_1,t_2$ and $G_{m1},G_{m2}:t_1=M_{11}G_{m1}+M_{21}G_{m2},t_2=M_{12}G_{m1}+M_{22}G_{m2}$. }
\label{SM_BZ}
\end{figure}
We perform the ED calculation on the finite momentum cluster, which can be expressed with $\bm{k}=n_1\bm{t}_1+n_2\bm{t}_2$, and $n_1,n_2$ are integers. Different geometries  yield different values for $\bm{t_1}$ and $\bm{t_2}$ as we show in Fig. \ref{SM_BZ}.

\subsection{The ED spectrum without self-consistent Hartree-Fock calculation}
\begin{figure}[h]
\includegraphics[width=\columnwidth]{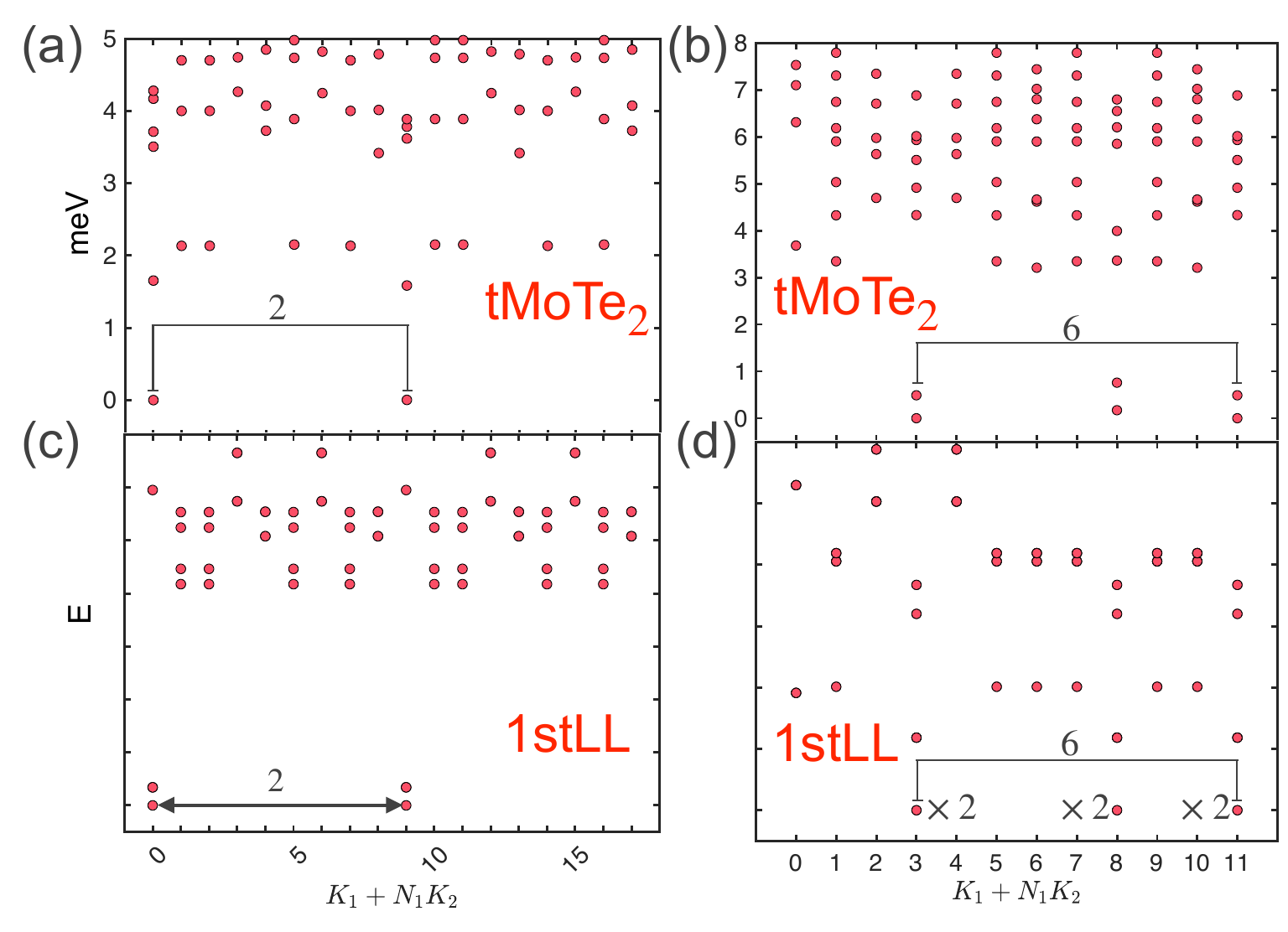}
\caption{Exact diagnoalization spectrum without SCHF. (a)/(b): the many-body spectrum from the ED calculation on the 18(12) clusterat half filling of 2nd moir\'e band with the continuum model of MoTe$_2$ at $\theta = 2.0^\circ$ and we use the dielectric constant $\epsilon = 5$. And we use periodic condition for the 18-site cluster and anti-periodic condition for 12-site cluster. (c)/(d):the same calculation on the 1st Landau Level, which agrees well with our results from continuum model. }
\label{SM_ED_NS_18}
\end{figure}

\begin{figure}[h]
\includegraphics[width=\columnwidth]{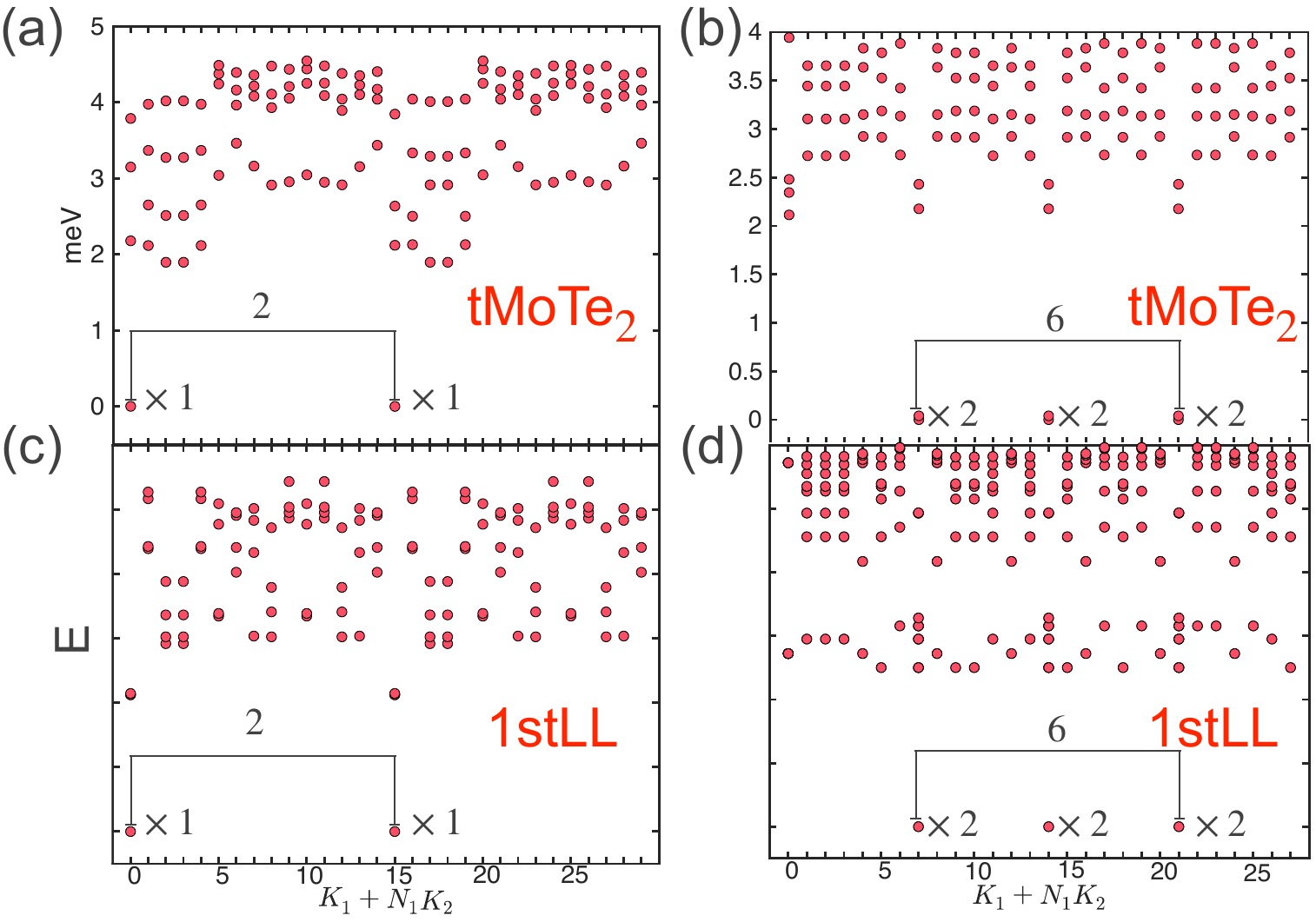}
\caption{Exact diagnoalization spectrum without SCHF. (a)/(b): the many-body spectrum from the ED calculation on the 30(28)-site cluster at half filling of 2nd moir\'e band with the continuum model of MoTe$_2$ at $\theta = 2.0^\circ$ and we use the dielectric constant $\epsilon = 5$.(c)/(d):the same calculation on the 1st Landau Level, which agrees well with our results from continuum model. }
\label{SM_ED_NS_30}
\end{figure}

In this part, we show the ED results(Figure. \ref{SM_ED_NS_30} and \ref{SM_ED_NS_18}) without the consideration of the effect on the filled 1st moir\'e bands, which displays six-fold quasi-degenerate states for even electrons on 12-site and 28-site clusters and two-fold quasi-degenerate states on 18-site and 30-site clusters. Both the momentum and degeneracy agree with the predictions of the generalized Pauli principle. And it also reveals remarkable similarities with the ground states of 1st-LL, as illustrated in Figs.~\ref{SM_ED_NS_18} and~\ref{SM_ED_NS_30}. And we use periodic condition($(\theta_1,\theta_2)=(0,0)$) for the 18-site cluster and anti-periodic($(\theta_1,\theta_2)=(\pi,0)$) condition for 12-site cluster. The twisted boundary condition is applied as:$\psi({\bm{r+L_i}})=e^{i\theta_i}\psi(\bm{r})$.


\subsection{The self-consistent Hartree-Fock calculation}
To account for the effect of the filled first moiré band, we perform a self-consistent Hartree-Fock calculation at $\nu = -2$. The general interaction Hamiltonian under Bloch basis reads:
\begin{equation}	H=\sum_{n\bm{k}\sigma}\epsilon_{n\bm{k}\sigma}a^\dagger_{n\bm{k}\sigma}a_{n\bm{k}\sigma}+\frac{1}{2}\sum_{\bm{k_1k_2k_3k_4}\sigma\sigma^\prime}^{n_1n_2n_3n_4}V_{\bm{k_1k_2k_3k_4}\sigma\sigma^\prime}^{n_1n_2n_3n_4}a^\dagger_{n_1\bm{k_1}\sigma}a^\dagger_{n_2\bm{k_2}\sigma^\prime}a_{n_3\bm{k_3}\sigma^\prime}a_{n_4\bm{k_4}\sigma}
\end{equation}
with momentum conservation:
\begin{equation}
\bm{q}	=\bm{k_1-k_4}=\bm{k_3-k_2}
\end{equation}
where n is the band index, $\sigma$ is the spin index, $\epsilon_{n\bm{k}\sigma}$ is the dispersion from the continuum model, $a^\dagger_{n\bm{k}\sigma}|0\rangle=|\psi_{n\bm{k}\sigma}\rangle$, $|\psi_{n\bm{k}\sigma}\rangle$ is the eigen-states of continuum model, and:
\begin{equation}
	\begin{aligned}
		V_{\bm{k_1k_2k_3k_4}\sigma\sigma^\prime}^{n_1n_2n_3n_4}
		&=\frac{1}{\Omega}\int \psi^*_{n_1\bm{k_1}\sigma}(\bm{r})\psi_{n_2\bm{k_2}\sigma^\prime}(\bm{r}^\prime)V(|\bm{r-r^\prime}|)\psi_{n_3\bm{k_3}\sigma^\prime}(\bm{r^\prime})\psi_{n_4\bm{k_4}\sigma}(\bm{r})d\bm{r}d\bm{r^\prime}\\		
\end{aligned}
\end{equation}
Apply the mean-field approximation on the interaction part:

\begin{equation}
	H_{\mathcal{MF}}=H_0+\sum_{\bm{k}\sigma\sigma^\prime}^{nn^\prime}h_{n\sigma,n^\prime\sigma^\prime}(\bm{k})a^\dagger_{n\bm{k}\sigma}a_{n^\prime\bm{k}\sigma^\prime}+E_c
		\end{equation}
with:
\begin{equation}
\begin{aligned}
\rho_{n_1n_2}^{\sigma\sigma^\prime}(\bm{k})&=\langle a^\dagger_{n_1\bm{k}\sigma}a_{n_2\bm{k}\sigma^\prime}\rangle\\
	h_{n\sigma,n^\prime\sigma^\prime}(\bm{k})&=	h^{hartree}_{n\sigma,n^\prime\sigma}(\bm{k})-	h^{fock}_{n\sigma,n^\prime\sigma^\prime}(\bm{k})\\
	h^{hartree}_{n\sigma,n^\prime\sigma}(\bm{k})&=	\frac{1}{2}\sum^{n_2n_3}_{\bm{k^\prime}\sigma^\prime}\rho^{\sigma^\prime\sigma^\prime}_{n_2n_3}(\bm{k^\prime})(V_{\bm{k^\prime kkk^\prime}\sigma^\prime\sigma}^{n_2nn^\prime n_3}+V^{nn_2n_3n^\prime}_{\bm{kk^\prime k^\prime k}\sigma\sigma^\prime})\\
	 h^{fock}_{n\sigma,n^\prime\sigma^\prime}(\bm{k})&=\frac{1}{2}\sum_{\bm{k^\prime}}^{n_2n_4}\rho_{n_2n_4}^{\sigma^\prime\sigma}(\bm{k^\prime})(V_{\bm{k^\prime kk^\prime k}\sigma^\prime\sigma}^{n_2nn_4n^\prime}+V^{nn_2n^\prime n_4}_{\bm{kk^\prime kk^\prime }\sigma\sigma^\prime})
\end{aligned}
\end{equation}
We then solve the mean-field Hamiltonian self-consistently. In our calculations, we select the converged states that preserve time-reversal symmetry. Throughout, we retain the top three moiré bands for both spins, to reduce compuatational cost. The mesh used in the Hartree-Fock calculations is typically nine times finer than that used in the ED calculations. For example, in the 30-site ED calculation, the Hartree-Fock mesh is 15 $\times$ 18. The ED calculation is performed on a sub-mesh of the converged mean-field states. 

\subsection{The extended data in other system size and many body Chern number}
\begin{figure}[h]
\includegraphics[width=\columnwidth]{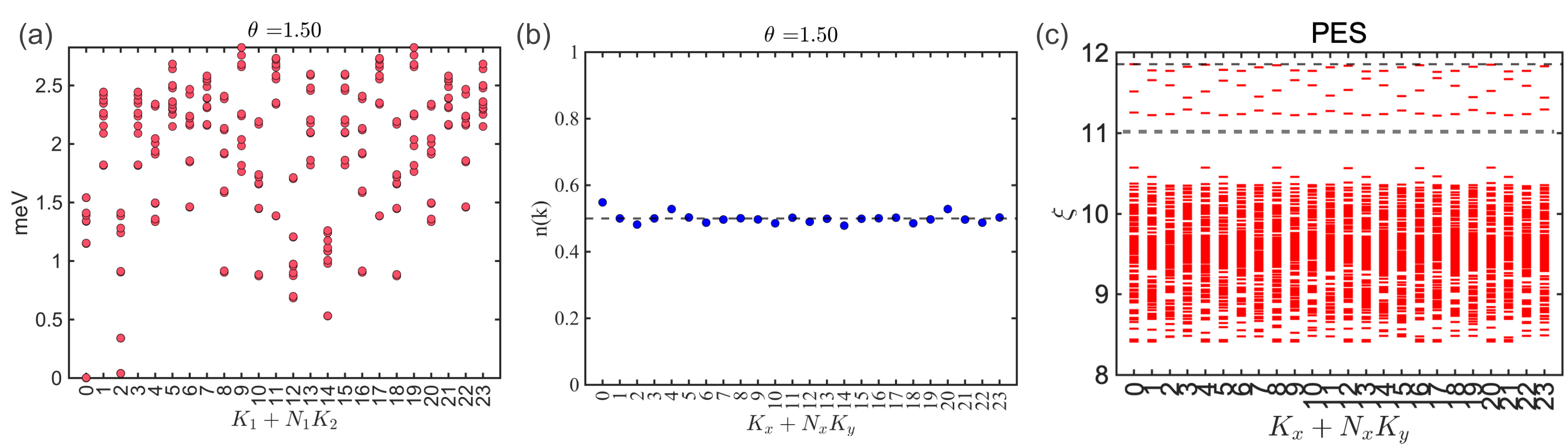}
\caption{(a). The ED spectrum for $4\times6$-site cluster with SCHF. The ground state manifold is six-fold degenerate at momenta k=0 and k=2. (b). The momentum distribution function $n(\bm{k})$.(c). The particle entanglement spectrum with a partition corresponding to $N_A = 3$ and there are 1952 states below the gap which matches with the generalized Pauli principle. }
\label{SM_ED_4x6}
\end{figure}

\begin{figure}[h]
\includegraphics[width=\columnwidth]{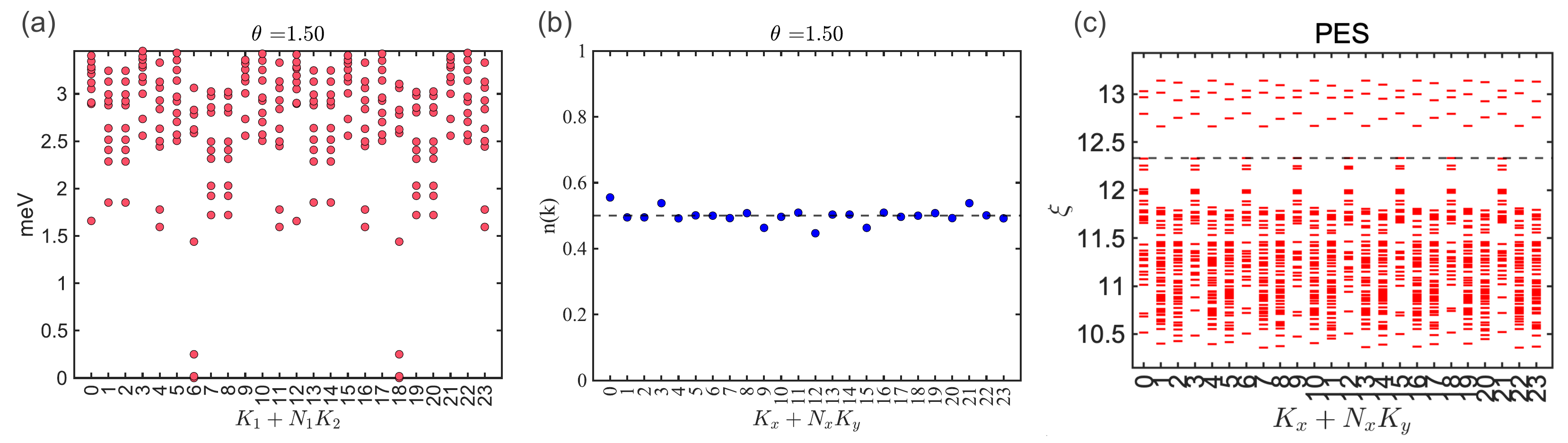}
\caption{(a). The ED spectrum for $3\times8$-site cluster with SCHF. The ground state manifold is six-fold degenerate at momenta k=6 and k=18. (b). The momentum distribution function $n(\bm{k})$.(c). The particle entanglement spectrum with a partition corresponding to $N_A = 3$ and there are 1952 states below the gap which matches with the generalized Pauli principle. }
\label{SM_ED_3x8}
\end{figure}

\begin{figure}[h]
\includegraphics[width=\columnwidth]{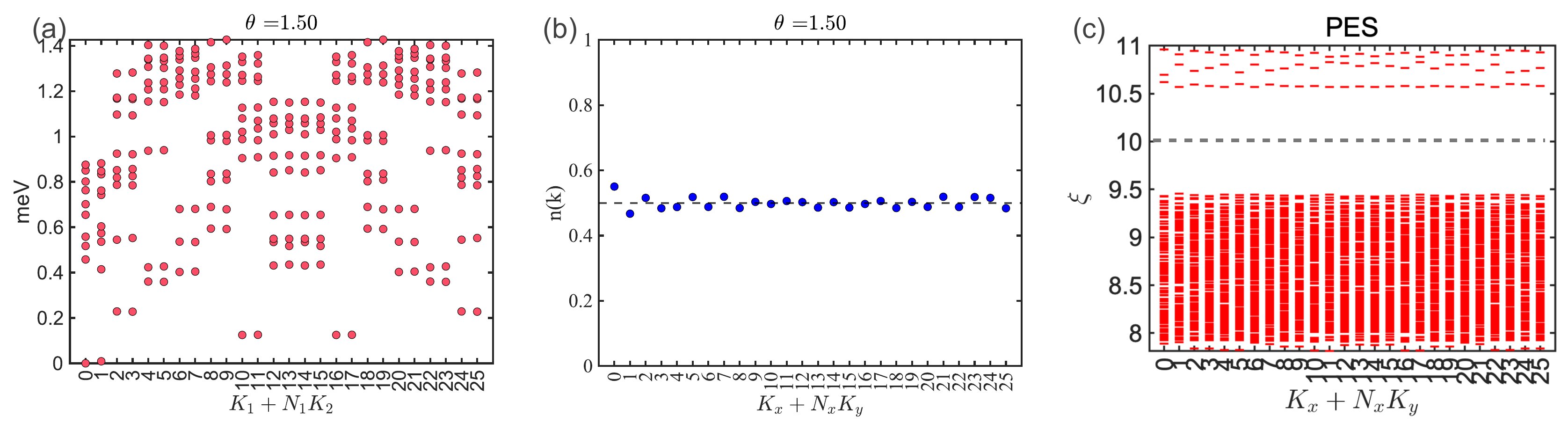}
\caption{(a). The ED spectrum for 26-site cluster with SCHF. The ground state manifold is two-fold degenerate at momenta k=0 and k=1. (b). The momentum distribution function $n(\bm{k})$.(c). The particle entanglement spectrum with a partition corresponding to $N_A = 3$ and there are 2522 states below the gap which matches with the generalized Pauli principle. }
\label{SM_ED_26}
\end{figure}

\begin{figure}[h]
\includegraphics[width=\columnwidth]{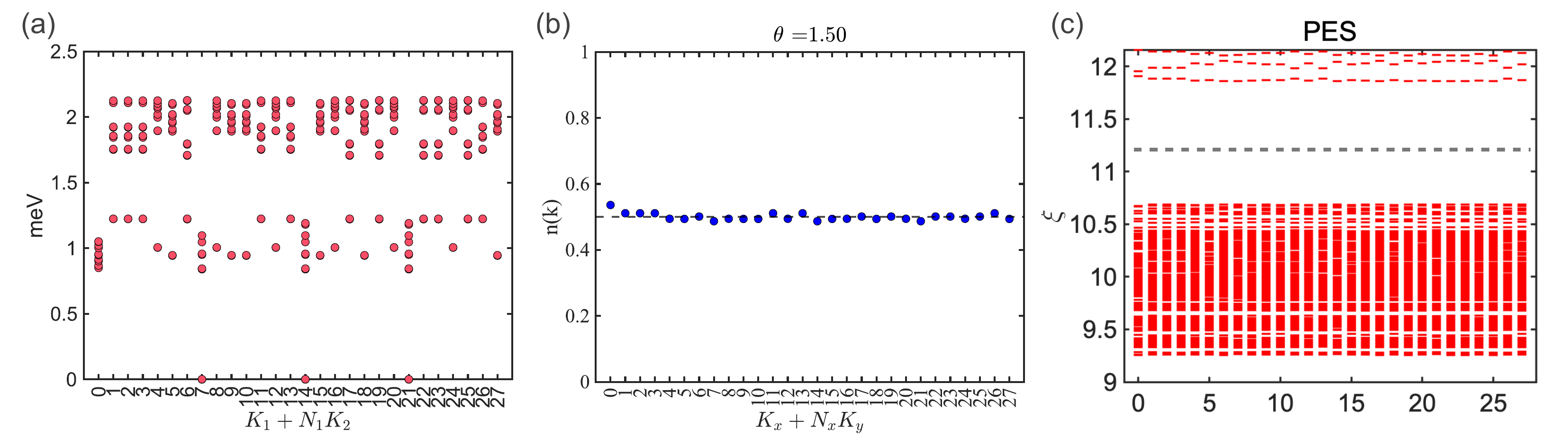}
\caption{(a). The ED spectrum for 28-site cluster with SCHF. The ground state manifold is six-fold degenerate at momenta k=7, k=14 and k=21. (b). The momentum distribution function $n(\bm{k})$. (c). The particle entanglement spectrum with a partition corresponding to $N_A = 3$ and there are 3192 states below the gap which matches with the generalized Pauli principle.}
\label{SM_ED_28}
\end{figure}

\begin{figure}[h]
\includegraphics[width=\columnwidth]{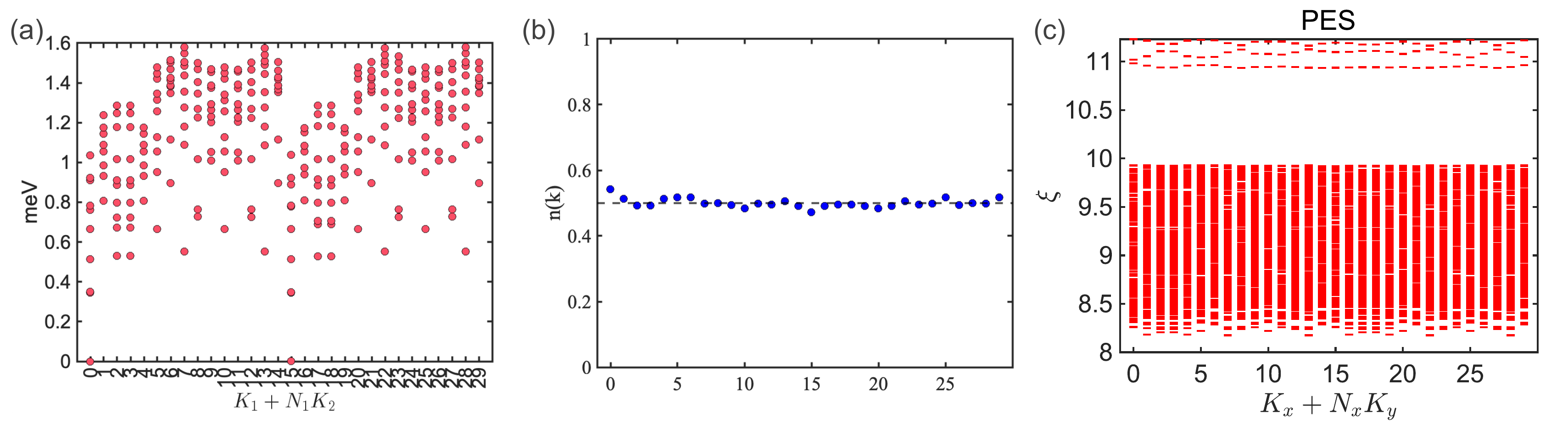}
\caption{(a). The ED spectrum for $5\times 6$-site cluster with SCHF. The ground state manifold is two-fold degenerate at momenta k=0 and k=15. (b). The momentum distribution function $n(\bm{k})$. (c). The particle entanglement spectrum with a partition corresponding to $N_A = 3$ and there are 3970 states below the gap which matches with the generalized Pauli principle. }
\label{SM_ED_30}
\end{figure}

In this section, we present all system sizes(Fig. \ref{SM_ED_4x6},\ref{SM_ED_3x8},\ref{SM_ED_26},\ref{SM_ED_28} and \ref{SM_ED_30})where the correct ground state degeneracy is obtained. We also show the momentum distribution function $n(\bm{k})$ and the particle entanglement spectrum (PES) with a partition corresponding to $N_A = 3$. Notably, $n(\bm{k})$ is highly uniform, and the states below the gap in the PES align with the quasi-hole excitations of the MR states.

\begin{figure}[h]
\includegraphics[width=\columnwidth]{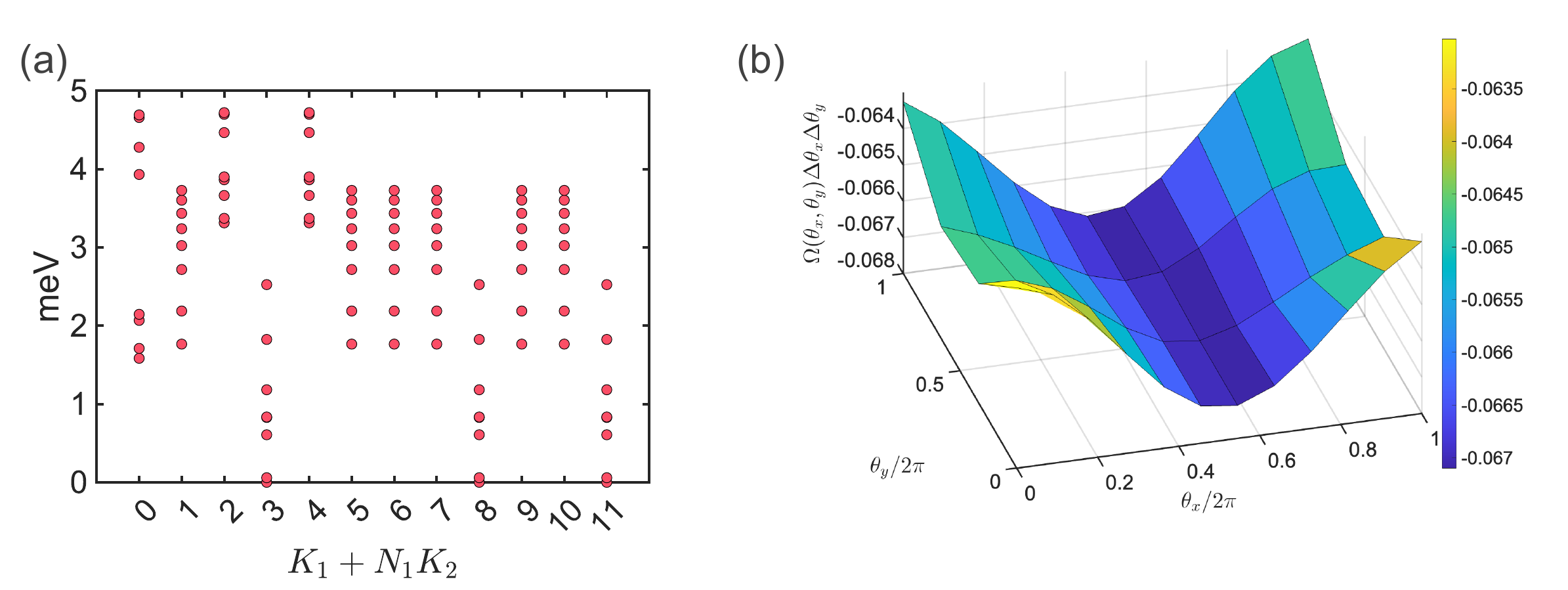}
\caption{ (a). The ED spectrum for 12-site cluster with SCHF. The ground state manifold is six-fold degenerate at momenta k=3, k=8 and k=11. (b). The total many-body Berry curvature distribution of the two ground states at momentum k=3. We note the integral of the total Berry curvature just equals to $C=\sum_{i=1}^2C_i=1$, namely $C_i=1/2$ per state.}
\label{SM_chern}
\end{figure}

Additionally, we present the many-body Berry curvature distribution for the 12-site cluster in Fig.~\ref{SM_chern} (b), which is quantized to 1, resulting in C=1/2 per state. For the 28-site system, we do not display the distribution due to the use of only a 3x3 mesh, but we still achieve perfect quantization. In the parameter plane of two independent twisted boundary angles $\theta_{x}\subseteq[0,2\pi],\theta_{y}\subseteq[0,2\pi]$, we can define the Chern number of the many-body ground state wavefunction $\psi(\theta_{x},\theta_{y})$ as an integral $C=\int\int d\theta_{x}d\theta_{y}\Omega(\theta_{x},\theta_{y})/2\pi$, with the Berry curvature $\Omega(\theta_{x},\theta_{y})=\mathbf{Im}\left(\langle{\frac{\partial\psi}{\partial\theta_x}}|{\frac{\partial\psi}{\partial\theta_y}}\rangle
-\langle{\frac{\partial\psi}{\partial\theta_y}}|{\frac{\partial\psi}{\partial\theta_x}}\rangle\right).$
Numerically, we divide the continuous parameter plane $(\theta_{x},\theta_{y})$ into $(m+1)\times(m+1)$ coarsely discretized mesh points $(\theta_{x},\theta_{y})=(2k\pi/m,2l\pi/m)$ where $0\leq k,l\leq m$. We can first define the Berry connection of the wavefunction between two neighboring points as $A_{k,l}^{\pm x}=\langle\psi(k,l)|\psi(k\pm1,l)\rangle$, $A_{k,l}^{\pm y}=\langle\psi(k,l)|\psi(k,l\pm1)\rangle$.
Then the Berry curvature on the small Wilson loop plaquette $(k,l)\rightarrow(k+1,l)\rightarrow(k+1,l+1)\rightarrow(k,l+1)\rightarrow(k,l)$ is given by the gauge-invariant expression $\Omega(\theta_{x},\theta_{y})\times4\pi^2/m^2=\mathbf{Im}\ln
\big[A_{k,l}^{x}A_{k+1,l}^{y}A_{k+1,l+1}^{-x}A_{k,l+1}^{-y}\big]$.
For a given ground state at momentum $K$, by numerically calculating the Berry curvatures using $m\times m$ mesh Wilson loop plaquette in the boundary phase space, we obtain the quantized topological invariant $C$ as a summation over these discretized Berry curvatures.

 
\subsection{The momentum counting from generalize Pauli principle}
\begin{figure}
\includegraphics[width=\columnwidth]{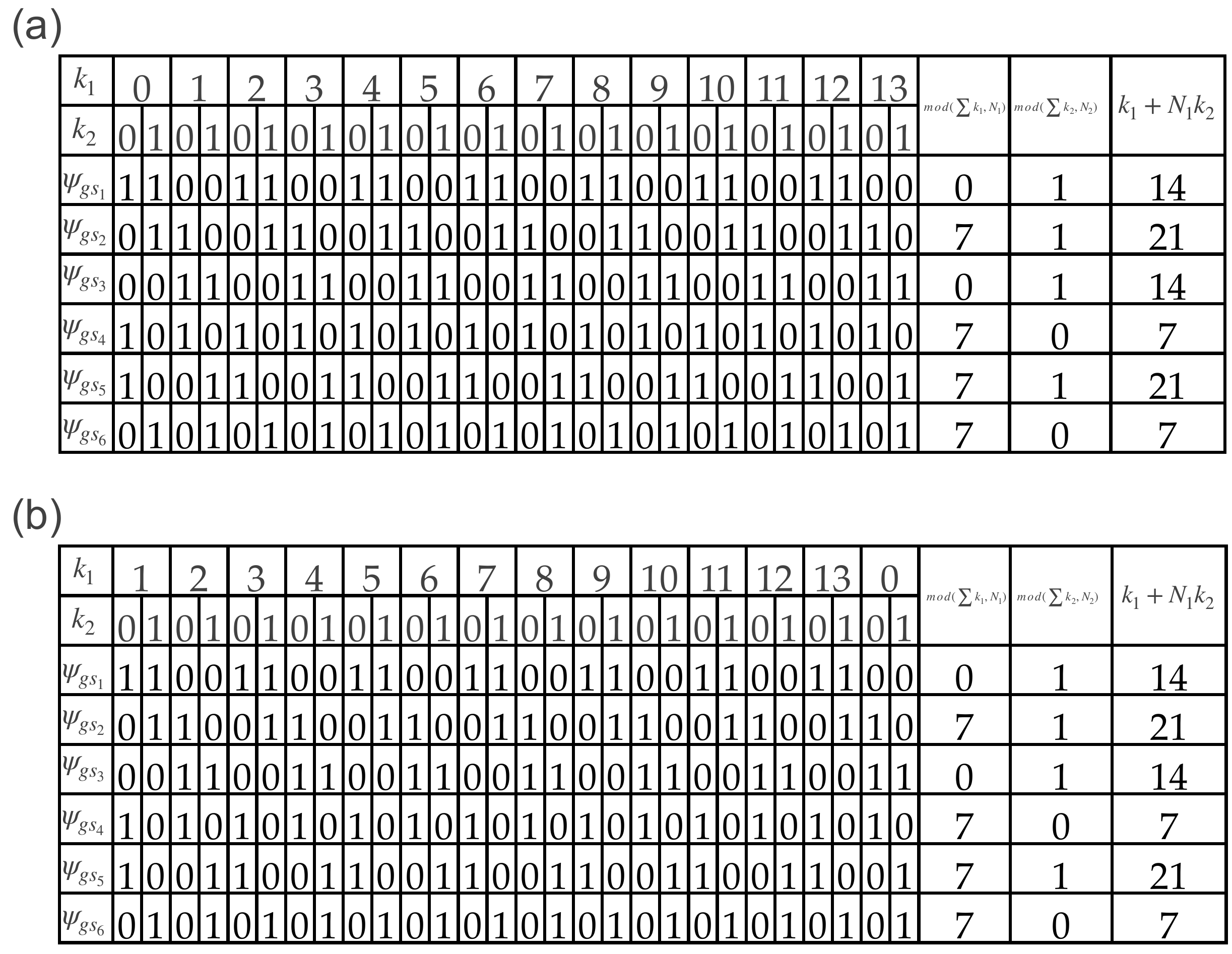}
\caption{The counting that obeys the (2,2) admissible rule for 28-site cluster. (a)The occupation partition when the twisted boundary condition reads:$(\theta_1,\theta_2)=(0,0)$. (b) The occupation partition when the twisted boundary condition reads:$(\theta_1,\theta_2)=(2\pi,0)$}
\label{SM_count28}
\end{figure}

\begin{figure}
\includegraphics[width=\columnwidth]{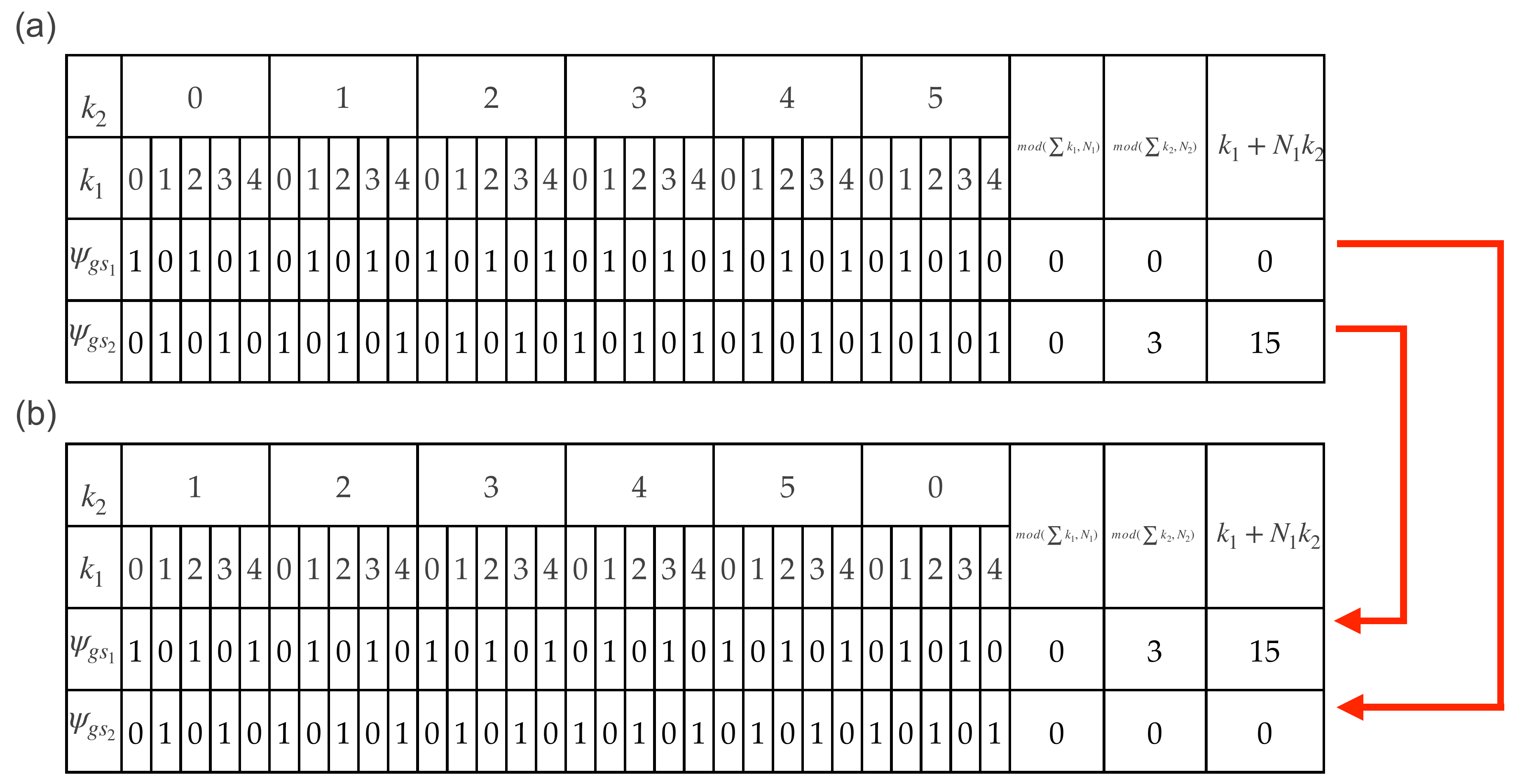}
\caption{The counting that obeys the (2,2) admissible rule for 30-site cluster. (a)The occupation partition when the twisted boundary condition reads:$(\theta_1,\theta_2)=(0,0)$. (b) The occupation partition when the twisted boundary condition reads:$(\theta_1,\theta_2)=(2\pi,0)$}
\label{SM_count30}
\end{figure}

With the generalized Pauli principle\cite{Haldane1991b,Bernevig2012} for Moore-Read states, which means there is no more than two electrons in four consecutive orbits, we calculate the momentum and degeneracy for the 28-site and 30-site clusters as shown in Fig.~\ref{SM_count28} and Fig.~\ref{SM_count30}. For 30 sites, only the occupation partition ``$1010\cdots101010$" and its translational invariant partner``$0101\cdots010101$" are allowed. And it also give the momenta 0 and 15 which agrees with our numerical results. For 28 sites, there are six possibilities denoted as $\Psi_{gs_1}$ through $\Psi_{gs_6}$. Additionally, the calculation of the total momentum yields three pairs of states, each with a two-fold degeneracy, all of which match with our exact diagonalization results. 

With the insertion of $2\pi$-flux quantum along the $t_2$ direction, the momentum shifts as $k_2 \rightarrow k_2 + 1$. The new occupation partition is shown in Fig.\ref{SM_count30}(b), where it is evident that $\psi_{gs_1}$ and $\psi_{gs_2}$ evolve into each other. However, for the 28-site cluster, as shown in Fig.\ref{SM_count28}, these ground states do not flow into each other.

\begin{figure}
\includegraphics[width=\columnwidth]{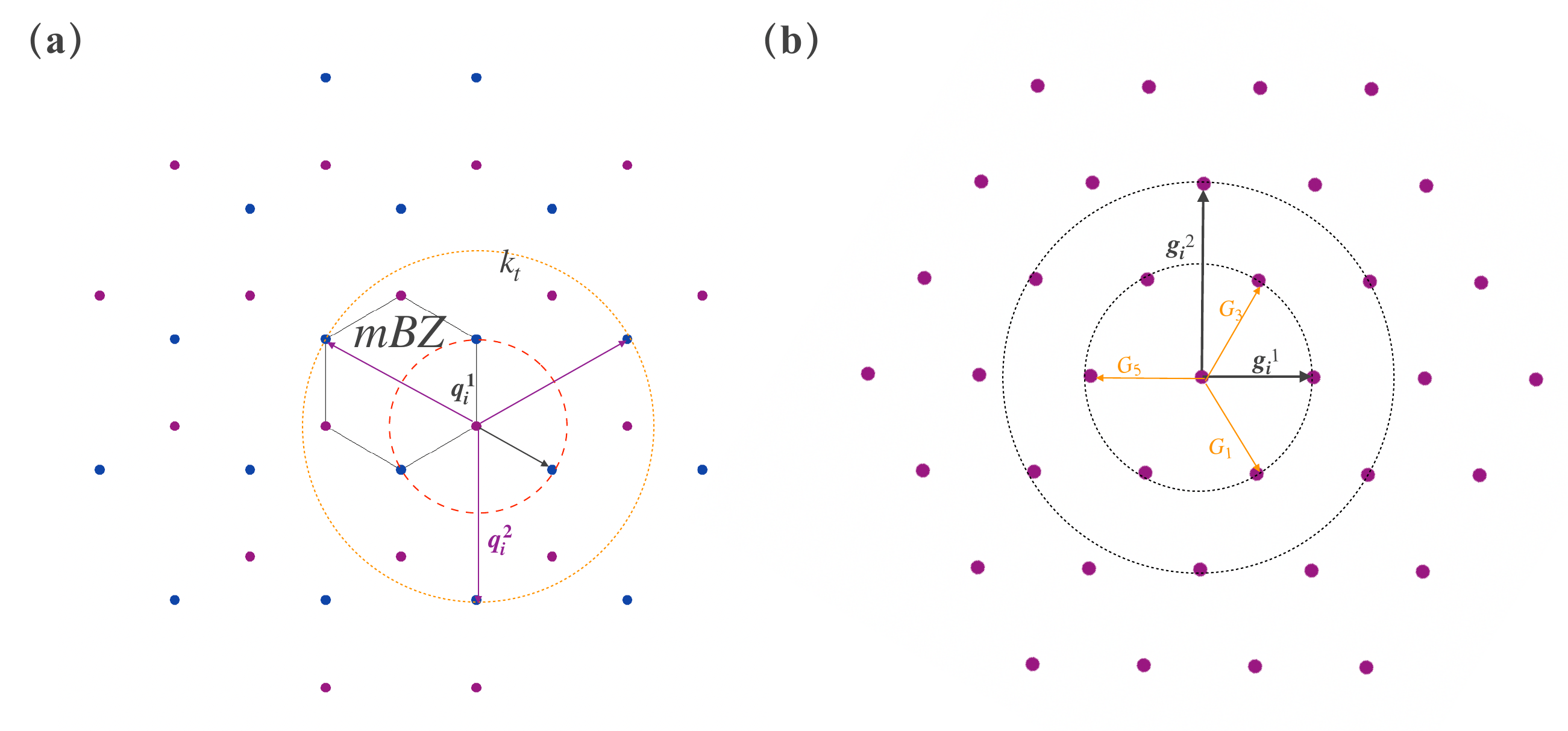}
\caption{
  (a) The purple dot signifies the plane wave of the top layer ($Q_{+}$), while the blue dot represents that of the bottom layer($Q_{-}$). The 1st and 2nd harmonic term related interlayer coupling momentum  $\bm{q^1_{i}}$ and $\bm{q^2_{i}}$ are illustrated here. 
  (b)  The 1st and 2nd harmonic term related intralayer coupling momentum  $\bm{g^1_{i}}$ and $\bm{g^2_{i}}$ are illustrated here. }
\label{SM_skyrmion}
\end{figure}
\subsection{The flux insert calculation on the Skyrmion model and tMoTe$_2$}
The Hamiltonian of Skyrmion model reads:
\begin{equation}
	\hat{H}=\frac{\hat{p}^2}{2m}+J\bm{ \sigma\cdot S(r)}
\end{equation}
And:
\begin{equation}
\begin{aligned}
&	\bm{S(\bm{r})}=\frac{\bm{N(r)}}{N(\bm{r})}\\	
&	\bm{N(r)}=\frac{1}{\sqrt{2}}\sum_{j=1}^6 e^{i\bm{q_j\cdot r}}\hat{e}_j+N_0\hat{z}\\
&    \hat{e}_j=(i\alpha \sin \theta_j,-i\alpha\cos\theta_j,-1)/\sqrt{2}\\
& \bm{q_j}=\frac{4\pi}{\sqrt{3}a_m}(\cos\theta_j,\sin\theta_j)
\end{aligned}	
\end{equation}
where the Skyrmion texture is determined by the parameter: $N_0 \quad \& \quad \alpha$

\begin{figure}
\includegraphics[width=\columnwidth]{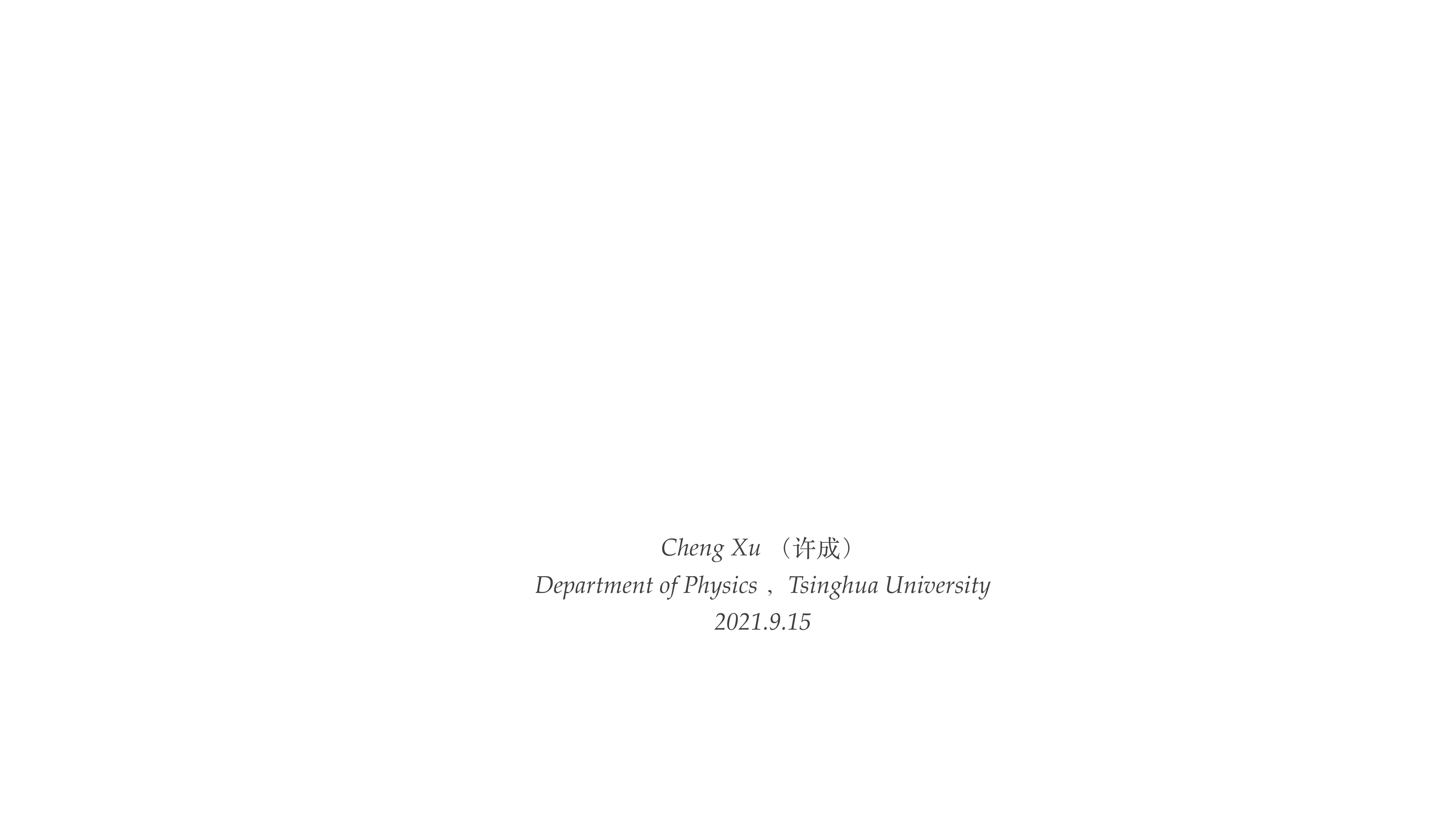}
\caption{(a)/(b) The spectrum flow of the Skyrmion model\cite{reddy2024non} on the 28/30-site cluster under the variation of twisted boundary condition. The quasi-degenerate states remain isolated with the excited states by the a visible gap. (c)/(d): the spectrum flow of the quasi-degenerate ground states on the 28/30-site cluster.}
\label{SM_flux}
\end{figure}

In detail:
\begin{equation}
	\begin{aligned}
&		\bm{N(r)}=(\frac{i}{2}\alpha\sum_j e^{i\bm{q_j\cdot r}}\sin\theta_j,-\frac{i}{2}\alpha\sum_j e^{i\bm{q_j\cdot r}}\cos\theta_j,N_0-\frac{1}{2}\sum_j e^{i\bm{q_j\cdot r}})\\
	\end{aligned}
\end{equation}
In flux calculation, we choose:
\begin{equation}
    \begin{aligned}
        m&=0.6m_e\\
        \alpha&=1\\
        N_0&=0.28   \\
        J&=0.5 eV\\
        a_m&=50\AA        
        \end{aligned}
\end{equation}
To further validate the behavior of MR states under flux insertion, we performed the same flux insert calculation in the Skyrmion model\cite{reddy2024non}(the filled first miniband is taken into account with the self energy.). Despite the differences in specific details, these states exhibit the same behavior under flux insertion: both the six-fold quasi-degenerate states and the two-fold quasi-degenerate
states maintain their degeneracy and are well separated from the excited states. Also only the two-fold degenerate states flow into each other and the six-degenerate states don't evolve into each other in Fig ~\ref{SM_flux}.
\begin{figure}[t]
\includegraphics[width=1\columnwidth]{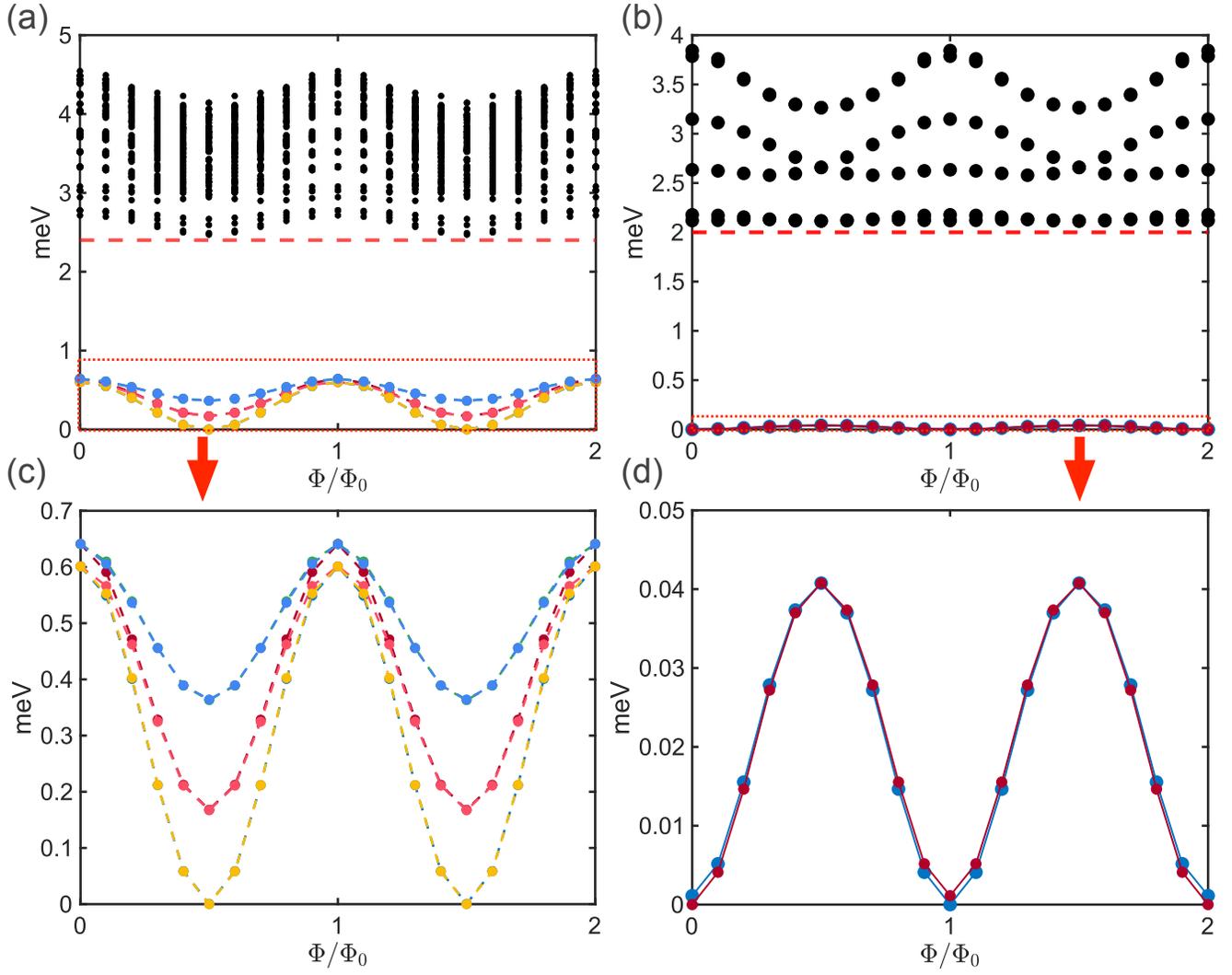}
\caption{Flux insert without SCHF(a)/(b) The spectrum flow of the twisted MoTe$_2$ on the 28/30 sites cluster under the variation of twisted boundary condition. The quasi-degenerate states remain isolated with the excited states by the a visible gap. (c)/(d) the spectrum flow of the quasi-degenerate ground states on the 28/30 sites cluster, which is also highlighted  in the (a)/(b) with the dashed red box.}
\label{NFig4}
\end{figure}
\begin{figure}[t]
 \includegraphics[width=1\columnwidth]{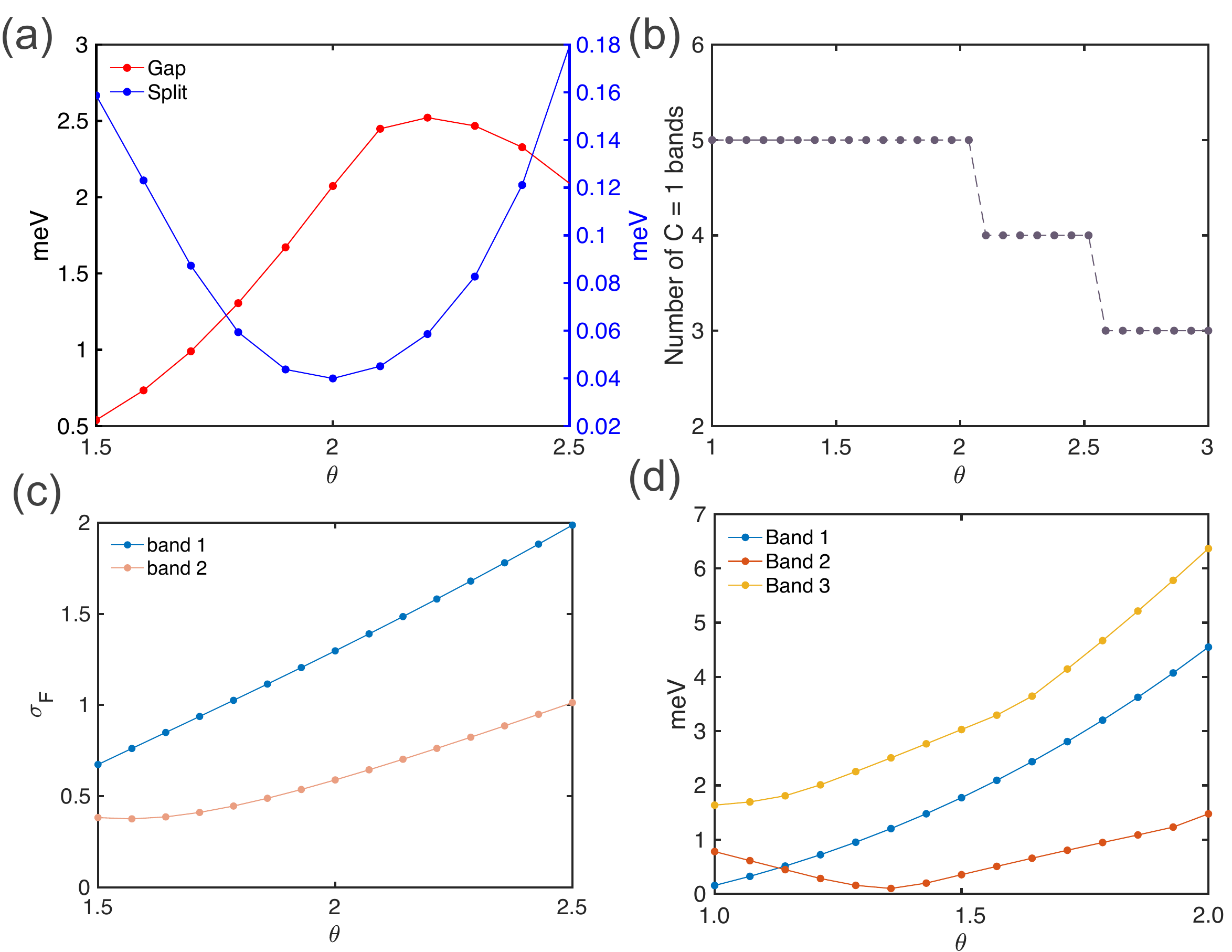}
 \caption{(a) The red dot denotes the band gap from the ED spectrum , which is defined as the energy difference between the quasi-degenerate states and the excited states. Besides, the blue dot is used to represent the energy spread between the quasi-degenerate ground states. Both data are derived from our ED calculation performed on the 28 sites cluster without SCHF. 
 (b) The number of bands which have chern number $C=1$ in the top five moir\'e bands from the continuum model, it agrees well with the local basis DFT results as we show in Fig.1(b) around 2.0$^\circ$(c) The fluctuation of Berry curvature of the top two band in the continuum model. (d) The bandwidth of top three moir\'e bands from the continuum model.
 }\label{Fig4}
 \end{figure}


\end{document}